%% file: main.tex
\documentclass[10pt]{iopart}

\usepackage{iopams}  
\expandafter\let\csname equation*\endcsname\relax

\expandafter\let\csname endequation*\endcsname\relax
\usepackage{amsmath}
\usepackage[version=4]{mhchem}
\usepackage{hyperref}
\usepackage{subcaption}
\usepackage{float}
\usepackage{graphicx}
\usepackage{color}
\usepackage{algorithm}
\usepackage{algpseudocode}
\graphicspath{{Plots}}
\newcommand{\cotwo}{$\mathrm{CO_2}\,$}

\newcommand{\revise}[1]{\textcolor{black}{#1}}

\begin{document}

\title[]{Triboplasma assisted chemical conversion in granular systems: a semi-analytic model}

\author{Hariswaran Sitaraman$^1$\footnote{Corresponding Author}, Sankaran Sundaresan $^2$}

\address{1. Computational Science Center, National Renewable Energy Laboratory, 15013 Denver W Pkwy, Golden, CO 80401, USA}
\address{2. Department of Chemical and Biological Engineering, Princeton University, Princeton, NJ, 08544, USA}
\ead{hariswaran.sitaraman@nrel.gov, sundar@princeton.edu}
\vspace{10pt}
\begin{indented}
\item[]September 2024
\end{indented}

\begin{abstract}
We present a semi-analytic model for a novel plasma-assisted chemical conversion pathway using triboplasmas generated in granular flows. Triboelectric charge relaxation is a well known phenomena where the potential generated from contact charging of particles exceeds the breakdown voltage of the background gas. In this work, we extend the triboelectric charge relaxation theory to include non equilibrium plasma energy and particle balance equations to predict the formation of dissociated and excited species that act as precursors to chemical conversion, for example in plasma-assisted ammonia synthesis. 
Our example case study with nitrogen background gas and teflon/aluminum tribomaterial system yielded high excited nitrogen species densities per collision that are comparable to current plasma-assisted conversion pathways. We also present a regime diagram for various gases where Paschen breakdown parameters are used to determine whether triboplasmas can be formed for a given effective work-function difference between two materials. Our sensitivity studies indicate particle velocity, particle radius, solids fraction and space charge effects play a critical role in overall plasma densities and excited species production.
\end{abstract}

%
\vspace{2pc}

\submitto{\JPD}
%
\maketitle
%
%
\section{Introduction}
\input{intro}
\section{Mathematical model}
\input{mathmodel}
%
\section{Results}
\input{results}

\section{Conclusions and future work}
\subsection{Conclusions}
\label{sec:conclusion}
In this work, we presented a semi-analytic model for producing excited and dissociated species in an \ce{N2} triboplasma from contact charging and discharge in a granular medium.  We developed a zero-dimensional plasma model using particle and material balances to extend the triboelectric charge relaxation model.  We developed three models with increasing complexity: an isolated particle with uniform velocity and charge, a homogeneous multi-particle system with uniform velocity, and a multi-particle system with a velocity distribution governed by a granular temperature. We used our models to study the formation of nitrogen excited and dissociated species when teflon particles collide with grounded aluminum walls. Significant species densities (N densities $\sim$ 1E17-1E18 $\mathrm{\#/m^3}$ and \ce{N2}(ex) densities $\sim$ 1E19-1E21 $\mathrm{\#/m^3}$) were obtained during charge relaxation that are comparable to plasma catalytic DBD systems (N densities $\sim$ 1E16-1E18 $\mathrm{\#/m^3}$ and \ce{N2}(ex) densities $\sim$ 1E21-1E23 $\mathrm{\#/m^3}$) for ammonia synthesis. Our model also predicted electron temperatures in the range of 1 eV that enables higher production of excited species compared to dissociated species. \ce{N2}(ex) density was about 1000 times higher than N density which improves energy efficiency of plasma catalytic ammonia synthesis. Our investigations also revealed that the presence of space charge with multiple particles led to weaker plasma densities ($\sim$ 3X lower N and \ce{N2}(ex) densities) when compared to a single particle triboplasma. The particle velocity, particle radius, number density, and reactor size were found to be the significant parameters that influence the yield of excited species. Specifically, larger particles, higher velocity, lower solids fraction, and thinner channel widths favor increased plasma densities per particle. We also presented a feasibility diagram for various gases where Paschen breakdown parameters were used to determine whether triboplasmas can be formed for a given work-function difference between two materials. It was found that gases like \ce{CO2} and \ce{SO2} were in the infeasible regime with the baseline material and particle parameters used in this work, while material choice with larger effective work function difference and geometric parameters (particle radius, channel width) can enable plasma relaxation in these high breakdown strength gases. Finally, extension of our model to granular systems showed that the overall excited species production is directly correlated to granular temperature and solids fraction.

This model is a first step towards investigating this promising pathway for plasma-assisted chemical conversion that is amenable to scale-up without up-scaling electrical power supply requirements. A tribo-plasma-based conversion process can be readily intensified using grounded internals to maximize wall-particle contacts. The number of microdischarges can also be controlled by changing particle number density and granular temperature. Our baseline granular case with a solids fraction of 0.01 and granular temperature of 100 $\mathrm{m^2/s^2}$ results in about 3 million microdischarges per second per square meter which is comparable to packed bed DBDs \cite{van2020plasma}. Tribo-plasma may also be achieved in fluidized beds with binary particles having different effective work functions (see Fig.\ \ref{fig:concept}), but it remains unexplored and unreported. Preliminary theoretical analysis to identify the formation of such microplasmas will be impactful, as a large number of inter-particle collisions per unit time can be readily achieved in binary fluidized beds.

\subsection{\revise{Future work}}

\revise{ This work evaluates the feasibility of a triboplasma assisted chemical conversion pathway using a model with simplified assumptions that include spherical particles, perfectly conducting surfaces, and zero dimensional plasma description, among others, and serves as a stepping stone for future research. One of our immediate efforts are along creating an experimental setup similar to a vibrating bed to test this theory, while improving our model to account various effects as listed below.}

\revise{Irregular particles can have interesting effects such as non-circular collisional overlap, electric field focusing at sharp corners, comminution, and charge exchange from material transfer. We plan to investigate these effects experimentally and simultaneously improve our model
accounting for these multidimensional aspects.}
\revise{Quantifying the conversion of mechanical energy during contact is essential for determining tribocharging efficiency, localized mechanochemistry, and material heating effects. These aspects will require thermo-mechanochemical models at the mesoscale that resolve interactions at contact, that will be later coupled to larger scale models via effective parameters such as tribocharging efficiency, heat transfer, and restitution coefficients. Moreover, the presence of moisture and Van-der-Waals cohesion/adhesion effects may affect tribocharging efficiency and effective restitution coefficients for smaller size particles (Geldart A/C) and lower velocities \cite{gu2016modified}. There is recent evidence as well on non-uniformity of charge distributions on the particle \cite{preud2023tribocharging} which affects the extent of the discharge. These aspects will explored experimentally as well as theoretically through modified collisional models within discrete-element-method based formulations. The effect of insulating walls as opposed to perfectly conducting walls used in this work presents additional challenges from surface charging along with modifications to image charge formulation in our work. We will utilize a numerical electrostatics solver as opposed to the simplified image charge formulation to account for these effects in our future efforts.}
\revise{The zero dimensional description used for our plasma model neglects the impact of plasma sheaths and secondary/field electron emission effects, that can have large impact on discharge behavior and breakdown strengths at small length scales. A 2 or 3D model for plasma dynamics along with detailed plasma chemistry is necessary to quantify these effects and is part of our ongoing efforts.}


\section{Supplementary material}
All codes and scripts used in this work are available as open-source at \url{https://github.com/hsitaram/triboplasma_codes}.


\section*{Acknowledgements}
This work was authored in part by the National Renewable Energy Laboratory, operated by Alliance for Sustainable Energy, LLC, for the U.S. Department of Energy (DOE) under Contract No. DE-AC36-08GO28308. Funding provided by U.S. Department of Energy's Laboratory Directed Research and Development (LDRD) is acknowledged. The views expressed in the article do not necessarily represent the views of the DOE or the U.S. Government. The U.S. Government retains and the publisher, by accepting the article for publication, acknowledges that the U.S. Government retains a nonexclusive, paid-up, irrevocable, worldwide license to publish or reproduce the published form of this work, or allow others to do so, for U.S. Government purposes.



\section*{References}
\bibliography{refs.bib}

\end{document}

%% file: intro.tex
\revise{Tribocharging (TC) or contact-charging refers to electrical charge exchange between two surfaces (can be same or different material) when they are in transient contact (e.g.,  impact) with each other \cite{matsusaka2010triboelectric,sotthewes2022triboelectric}.} TC is a well-known phenomena in the context of granular flows and has been utilized in an advantageous way in many particulate applications such as electrophotography \cite{pai1993physics,schein1999recent}, powder coating \cite{kopp2023enabling}, and air filtration \cite{wang2021tribo}. Recently, TC has been studied as a way to harvest energy using  triboelectric nanogenerators (TENG) \cite{wang2020triboelectric,cheng2023triboelectric} for sensors and microelectronics applications. \revise{TC is also being investigated as a way to form self-assembled macroscopic crystals with favorable mechanical and optical properties for photonics, material processing, and electronics applications \cite{battat2023melting, grzybowski2003electrostatic, fendler2001chemical,sotthewes2024toward}.} Often times TC has a negative impact on industrial processes when electrostatic interactions result in particle clustering, reduced mixing efficiencies, and even lead to shutdown of reactors from discharge-driven explosions \cite{liu2020effect}.

One of the important physical phenomena that happens frequently after contact electrification is the relaxation of charge attained by the particle via dielectric breakdown of the gaseous medium surrounding the particle. Matsuyama and Yamamoto \cite{matsuyama1995charge} developed a  theoretical model for charge relaxation by considering the extent to which the potential generated by the contact charge achieves the breakdown voltage governed by the Paschen curve. This relaxation model continues to be used today in complex granular flow models using discrete element methods (DEM) \cite{liu2022effect,wang2023cfd} where studies are mostly tailored towards understanding and mitigating granular flow upsets in industrial systems.  In this paper, we recognize that this dielectric breakdown phenomenon in TC is a non-thermal plasma discharge and ask the question if such a plasma can be used in a beneficial way for plasma-assisted chemical conversion. 

Several recent studies with non-thermal plasmas using sources such as dielectric barrier discharges (DBD) \cite{li2019review} have shown promise towards sustainable chemical production using renewable electricity. Ammonia generation/decomposition \cite{mehta2020plasma,chen2021plasma,wang2022shielding}, and \cotwo conversion  to clean fuels and chemicals \cite{bogaerts2020plasma} are some of the many pathways that have been recently investigated using non-thermal plasmas. Recently, TENGs have also been used as an independent device to power a plasma discharge for \cotwo conversion achieving mechanical energy conversion efficiency of 2.3\% \cite{li2022triboelectric}. There exists several outstanding challenges associated with scale-up and energy-efficiency for non-thermal plasma-assisted conversion \cite{delikonstantis2022low}. Up-scaling high voltage electrical power supply for atmospheric pressure plasma operation is one of the limiting factors.  Can we circumvent this scaling problem by eliminating the external power and creating electrically charged particles and several discharges in-operando through TC?
\begin{figure}
    \centering
    \includegraphics[width=0.5\linewidth]{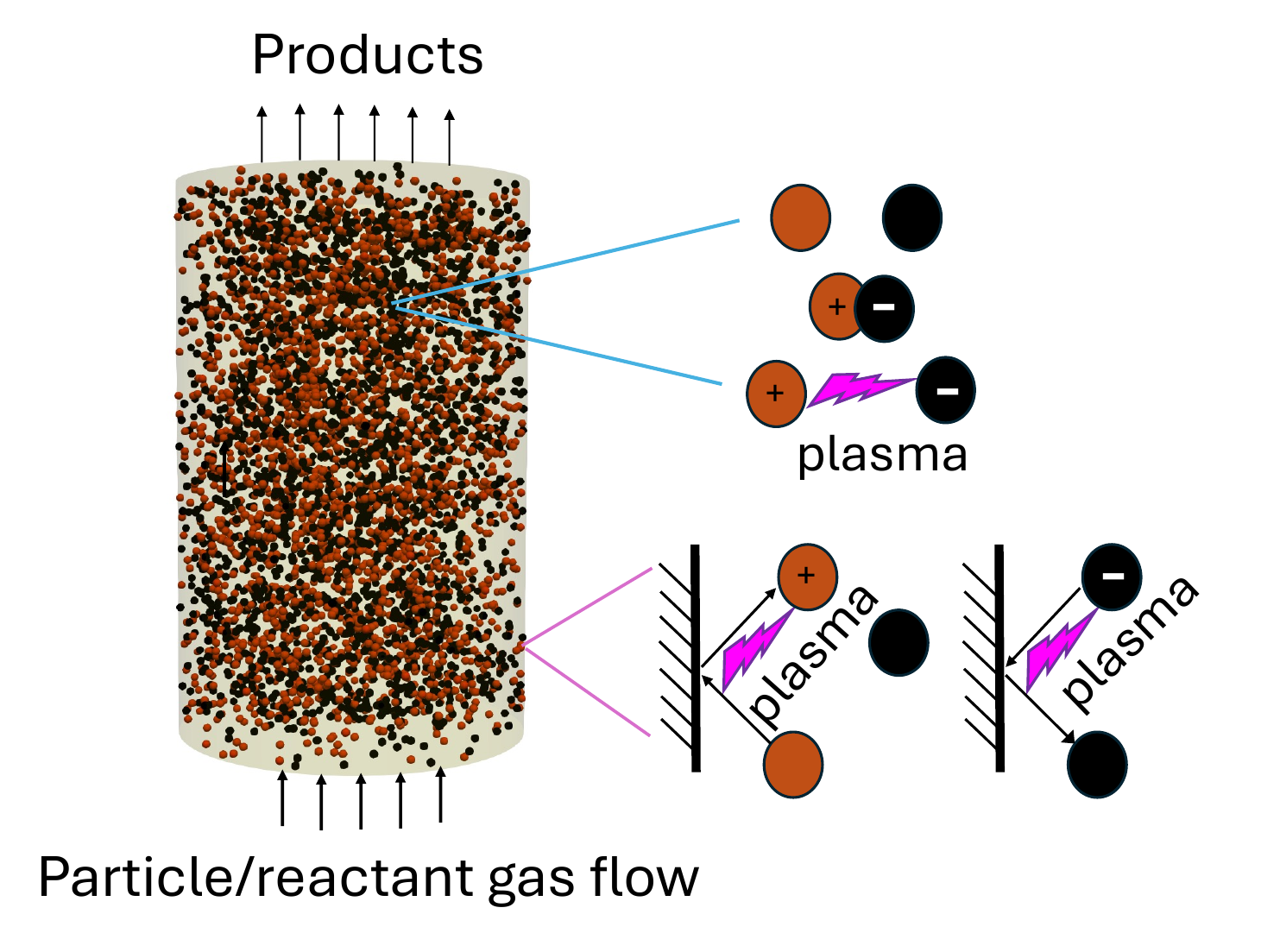}
    \caption{A conceptual contact charging assisted conversion reactor with catalyst (black) and carrier (orange) particles along with different plasma generation mechanisms that include plasma generation during carrier-catalyst, carrier-wall, and catalyst-wall interactions.}
    \label{fig:concept}
\end{figure}

Tribocharged particles are well known for producing intense electrical discharges. Examples from nature include ice crystal-graupel interactions \cite{sun2021aerosol} or volcanic ash charging \cite{cimarelli2022review}, both leading to terrestrial lightning events. A system can now be conceived where the kinetic energy of particles is utilized for contact charging which in turn results in non-equilibrium plasmas. In this way, a plasma environment with excited/dissociated gas-phase species can be realized. The chemical conversion yield can be further increased using plasma catalytic synergies \cite{bogaerts20202020} with the inclusion of catalytic surfaces or particles. Fig.\ \ref{fig:concept} shows a conceptual system where the granular phase includes both catalyst (black) and carrier (orange) particles. The carrier particles in Fig.\ \ref{fig:concept} can be conceived as easily tribochargeable particles with comparatively lower or higher effective work functions compared to the catalyst or reactor walls. Contact charging happens near the wall as well as during carrier-catalyst particle collisions and subsequent charge relaxation leads to several localized discharges. Several of these small scale discharges provide a reactive chemical environment with excited and dissociated species near catalyst particles, for conversion of gas mixtures such as $\mathrm{N_2}$ and $\mathrm{H_2}$ to ammonia. Multiple advantages to such a system can be hypothesized. Firstly, it eliminates the need for an external electrical power supply which is a significant bottleneck for scaling up plasma-catalytic conversion processes. Secondly, albeit the large system sizes, discharges at relatively smaller length scales ($\sim$ 10-1000 $\mu$m) (microdischarges) will be the driving plasma phenomena which are found to be superior in terms of current, energy and reactive species densities \cite{kushner2005modelling, sitaraman2011gas}. Thirdly, triboplasmas are produced in the vicinity of catalyst particles or surfaces thus reducing the chance of gas phase recombination or other detrimental pathways prior to catalytic activation. Particle size and density can be controlled to provide high surface-to-volume ratios for improved conversion. Finally, microdischarges are known to produce intense localized gas heating that can enable thermocatalytic synergies alongside plasma catalysis \cite{kushner2005modelling,sitaraman2011gas}.

The aforementioned advantages regarding such a novel pathway requires resolution of 
several fundamental scientific questions before this concept can be translated into a realistic system. In this paper, we make a first step towards developing a mathematical model for charge relaxation and subsequent plasma generation during separation. We study nitrogen discharges in this work motivated by significant recent interests in plasma-assisted ammonia generation. Our model quantifies the feasibility of conversion pathways and determine important granular flow and contact electrification parameters that determine overall yield of our conceived novel system.

%% file: mathmodel.tex
\label{sec:math_model}
\begin{figure}
    \centering
    \includegraphics[width=0.8\linewidth]{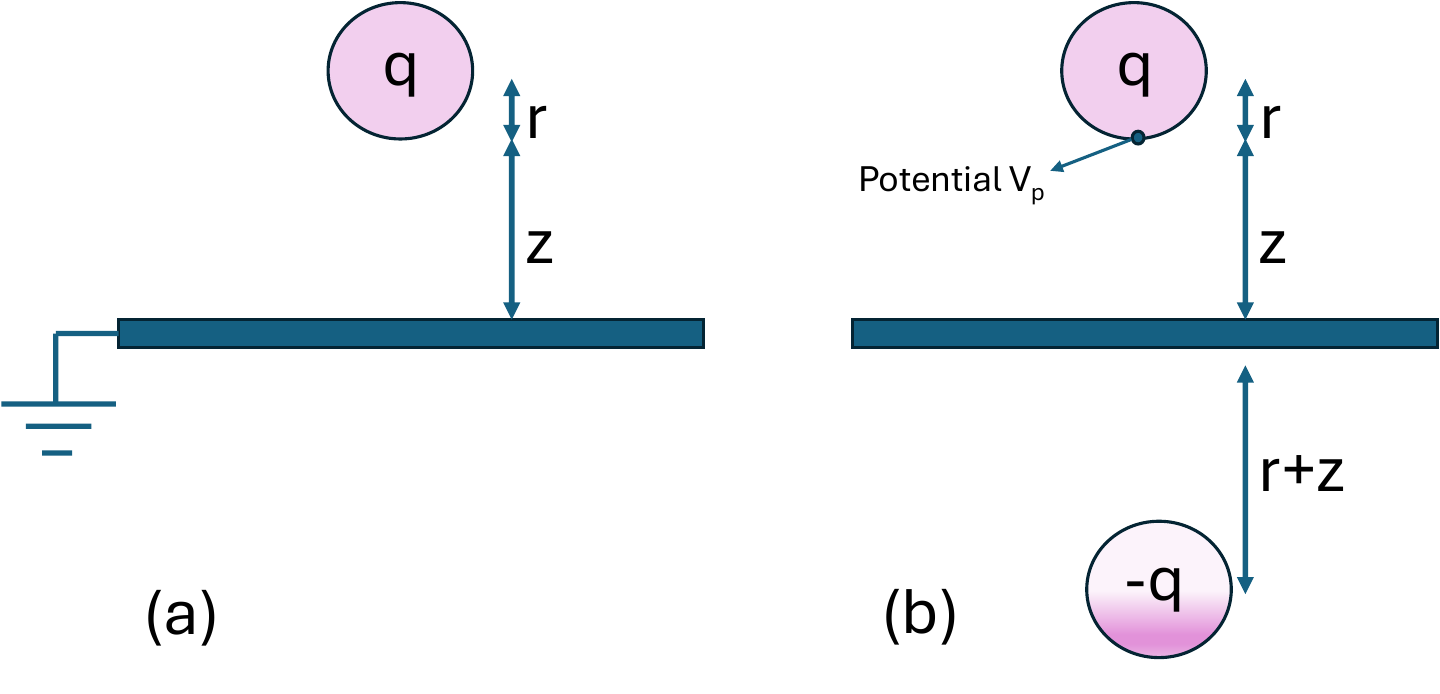}
    \caption{(a) \revise{polymer} particle (Teflon in this work) with charge $q$ above a grounded \revise{conducting} plate (Al in this work) and (b) image charge formulation to achieve zero potential at the conducting plate.}
    \label{fig:imgcharge}
\end{figure}
In order to quantify the plasma parameters (e.g. electron density, energy) achieved during triboelectric charge relaxation, a mathematical model that couples charged particle potentials, contact charging, gas breakdown, and plasma chemistry is necessary.
We first briefly review the derivation of the charge relaxation model by Matsuyama and Yamamoto \cite{matsuyama1995charge} that quantifies the propensity for gas breakdown from voltage generated by charged particles and using the condenser model \cite{matsusaka2010triboelectric} for contact charge exchange. We then incorporate plasma energy and particle balance equations that couple with the aforementioned charge relaxation model along with plasma chemistry for a nitrogen discharge.
\subsection{Charge relaxation model}
Let us first consider a single \revise{spherical polymer} particle colliding with a grounded \revise{conducting} wall as shown in Fig.\ \ref{fig:imgcharge}. The charge relaxation process involves, first the approach of the particle with a velocity $v_i$ and charge $q$, then triboelectric charge exchange at contact, and subsequent separation. At some distance from the wall during separation, gas breakdown is initiated if the critical breakdown potential is achieved, and the charge on the particle reduces as it recedes away from the wall. The relaxation continues until the reduced charge on the particle is unable to sustain a discharge. 

In order to quantify this relaxation process, the potential difference between the charged particle and the grounded wall using the image charge formulation is first derived, as was outlined in the review by Matsusaka et al. \cite{matsusaka2010triboelectric}. 
Consider a particle with radius $r$ and charge $q$ at a distance $z$ from the plate as shown in Fig.\ \ref{fig:imgcharge}. The potential difference $V_{img}$ between the particle and the plate can be obtained by placing an image charge equal in magnitude and opposite in polarity so as to achieve null potential at the grounded plate:
\begin{align}
    V_{img}(q,z) = \frac{1}{4\pi \epsilon_0}\left(\frac{q}{r}-\frac{q}{r+2z}\right)\\
    V_{img}(q,z) = \frac{q z}{2 \pi \epsilon_0 r (r+2z)} \label{eq:imcharge}
\end{align}
The final charge after collision using the condenser model \cite{matsusaka2010triboelectric} is given by: 
\begin{align}
    q_{f}=q+\delta q\\
    \delta q=k_c \frac{\epsilon_0 A_{coll}}{\delta_c} \left(V_c - V_{img}(q,\delta_c)\right) \label{eq:tribocharge}
\end{align}
where $\delta_c$ is the critical separation distance, $k_c$ is the contact charging efficiency and, $V_c$ is the effective work function difference between particle and wall materials. $A_{coll}$ is the area at contact, that depends on particle diameter and incident velocity, from the Hertzian collision model \cite{matsusaka2000electrification}:
\begin{align}
    A_{coll}=1.36 (k \rho_p)^{2/5} 4 r^2 v_i^{4/5}\\
    k=\frac{1-\nu_p^2}{E_p}+\frac{1-\nu_w^2}{E_w}
\end{align}
 For simplicity in the derivation at this point, we assume there are no space charge effects from multiple particles and that there are no external voltages applied in this system. The effect of space charge from multiple particles and the impact of velocity distributions are considered in upcoming sections \ref{sec:collectpart1} and \ref{sec:granular}, respectively. After collision, particle traveling with velocity $v_i$, rebounds with velocity $v_f=e v_i$, where $e$ is the coefficient of restitution. 
\begin{figure}
    \centering
    \includegraphics[width=\linewidth]{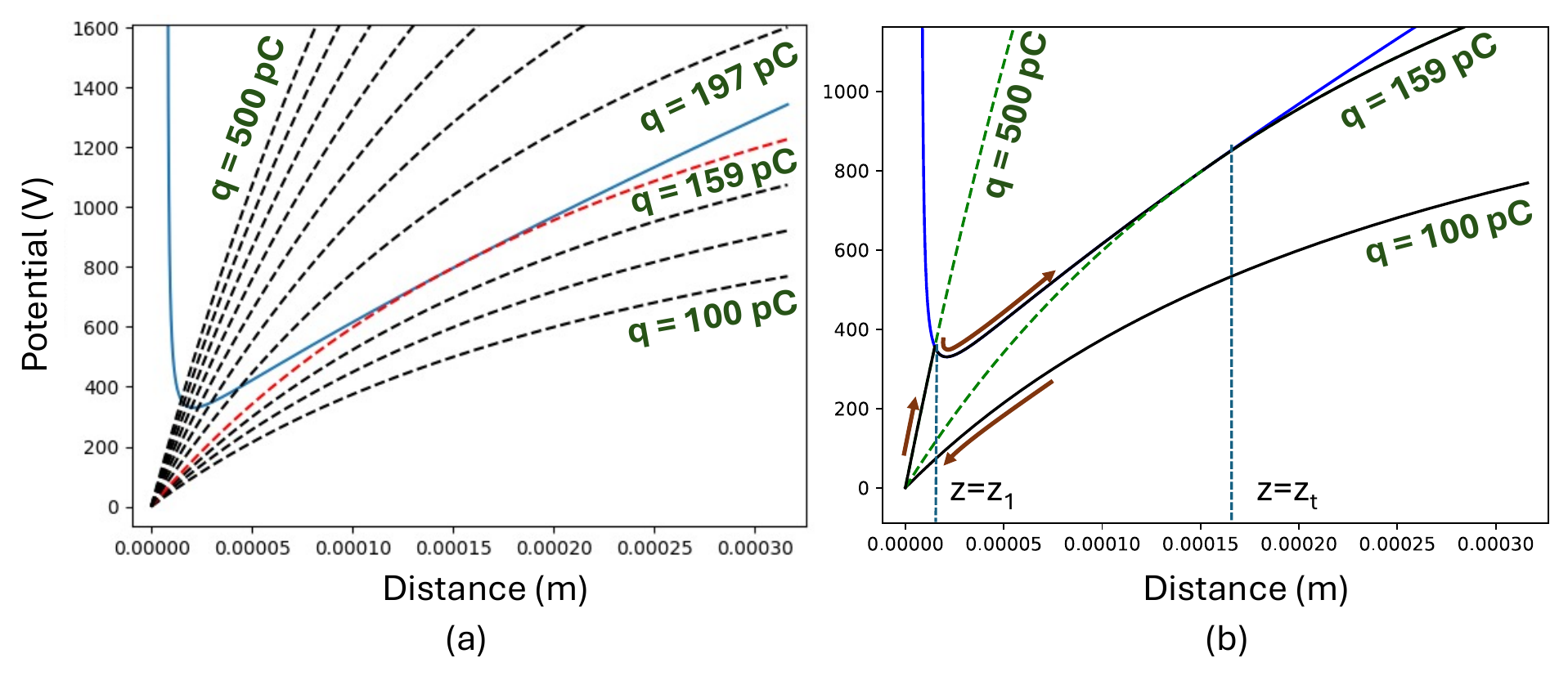}
    \caption{(a) An example Paschen curve with minimum voltage of 330 V and a minimum pressure times gap distance $pd_{min}=\mathrm{15.9 mm\,Torr}$ along with image-charge based potential curves (particle radius $r=\mathrm{0.6\,mm}$) with the red curve indicating the tangent case  and (b) shows a hypothetical example where an incoming particle with charge of 100 pC, collides and gains a post collision charge of 500 pC, whose potential curve after collision intersects with the Paschen curve at $z=z_1$, relaxes in charge between $z_1$ and $z_t$ and leaves along the tangent image potential curve at $z=z_t$. \revise{The Paschen curve (shown at atmospheric pressure) and particle radius used in this figure are hypothetical and merely used to explain the charge relaxation theory.}}
    \label{(fig:paschen_img_pot}
\end{figure}

The post contact charge relaxation can now be derived using the particle-wall potential difference and the increment in its charge from tribocharging. The strategy here is to compare the charged particle potential against the critical breakdown voltage of the gas as a function of separation distance to determine discharge initiation and termination.

Fig.\ \ref{(fig:paschen_img_pot} (a) shows an example Paschen curve at atmospheric pressure with a family of image potential curves with varying particle charge as a function of distance from the grounded wall. It is seen that the image charge potential curve intersects the Paschen curve if the particle has sufficient charge. An example hypothetical scenario shown in Fig.\ \ref{(fig:paschen_img_pot}(b) shows a particle with an incoming charge (100 pC in this case) that gains enough contact charge (400 pC)  to intersect the Paschen curve ($B_p(z)$) at the charge relaxation onset distance, $z=z_1$. $z_1$ can be obtained by equating image and Paschen curve potentials, given by:
\begin{align}
    \frac{q_{f} z}{2 \pi \epsilon_0 r (r+2z)}=B_p(z) \label{eq:ps_img}\\
    B_p(z)= V_{min} \frac{\left(\frac{pz}{pd_{min}}\right)}{1+\ln\left(\frac{pz}{pd_{min}}\right)} \label{eq:Paschen} \\
    B_p(z) = \frac{B (pz)}{C+ln(pz)} \label{eq:PaschenBC}\\
    B=\left(\frac{V_{min}}{pd_{min}}\right) \label{eq:PaschenB}\\
    C=\left(1.0-\ln(pd_{min})\right) \label{eq:PaschenC}
\end{align}
The parameters for the Paschen curve include the minimum breakdown voltage $V_{min}$, pressure $p$, and the critical $\mathrm{pressure \times electrode\,gap}$, $pd_{min}$, where the breakdown voltage is minimum. The parameters B and C in Eq.\ \ref{eq:PaschenBC} are obtained from Eq.\ \ref{eq:Paschen} after simplifying the $pd_{min}$ and $V_{min}$ terms, giving rise to Eqs.\ \ref{eq:PaschenB} and \ref{eq:PaschenC}, respectively.

The voltage between the particle and the plate then follows the Paschen curve as shown in Fig.\ \ref{(fig:paschen_img_pot}(b) between $z_1$ and $z_t$. The charge relaxation model assumes that every point along the Paschen curve corresponds to an image potential curve (Fig.\ \ref{(fig:paschen_img_pot}(a)) of a given relaxed charge until the particle potential curve is tangential to the Paschen curve at $z=z_t$ as shown in Fig.\ \ref{(fig:paschen_img_pot} (b). $z_t$ is where the particle would leave the Paschen breakdown curve and retains the final relaxed tangential charge ($q_{t}$) as it moves away from the wall.  Therefore, to get $z_t$ and $q_{t}$, we have two constraints: 1) the matching of slopes and 2) the matching of voltage between image potential curve and Paschen curve, which results in a non-linear equation system, given by:
\begin{align}
    \frac{dV_{img}}{dz}(q_{t},z_t) = \frac{d B_{p}}{dz}(z_t) \label{eq:tangent1}\\
    V_{img}(q_{t},z_t)=B_p(z_t) \label{eq:tangent2}
\end{align}
\subsection{Plasma model coupling with charge relaxation}

In this section, we couple a zero-dimensional plasma model \cite{lieberman1994principles} with charge relaxation parameters obtained from the previous section.
We can assume that the plasma is active between $z=z_1$ and $z=z_t$ over a 
time $\delta t=\frac{z_t-z_1}{v_f}$. Let us use a zero-dimensional description 
of the plasma with the aim of obtaining two important parameters, electron density $n_e$ and 
electron temperature $T_e$, that will close the chemical rate equations. These two parameters are
obtained using particle balance and energy balance constraints, as described below. This analytical strategy is similar to that described by Lieberman and Lichtenberg \cite{lieberman1994principles} for direct-current and capacitive discharges.
%
We consider the breakdown of \ce{N2} in this work because of well documented Paschen curve data and its recent importance on plasma catalytic ammonia synthesis. Consider a simple reaction scheme for nitrogen plasma with mass action kinetics and Arrhenius reaction rate constants for ionization, dissociation and excitation reactions
as shown below:
\begin{align}
    \ce{N2} + e \rightarrow \ce{N2}^+ + 2e \,,\Delta H=E_{iz}\nonumber\\
    R_{iz}=k_{iz}(T_e) n_e N_G \nonumber\\
    k_{iz}(T_e)=A_{iz} T_e ^{\alpha_{iz}} \exp(-T^{iz}_{a}/T_e)\\
    \ce{N2} + e \rightarrow \mathrm{N} + \mathrm{N} + e \,,\Delta H=E_{d}\nonumber \\
    R_{d}=k_{d}(T_e) n_e N_G \nonumber\\
    k_d(T_e)=A_{d} T_e^{\alpha_d} \exp(-T^{d}_{a}/T_e)\\
    \ce{N2} + e \rightarrow \mathrm{N_2(ex)} + e \,,\Delta H=E_{ex} \nonumber \\
    R_{d}=k_{ex}(T_e) n_e N_G \nonumber\\
    k_{ex}(T_e)=A_{ex} T_e^{\alpha_{ex}} \exp(-T^{ex}_{a}/T_e)
\end{align}
Here, the enthalpy of reaction $\Delta H$, is the ionization ($E_{iz}$), dissociation ($E_d$), and excitation ($E_{ex}$) energies, respectively corresponding to inelastic electron energy loss per collision with background \ce{N2}. $n_e$ and $N_G$ correspond to the number densities of electrons and background \ce{N2} gas, respectively.
The Arrhenius rate parameters for ionization ($A_{iz}$, $\alpha_{iz}$ and $T_a^{iz}$), dissociation ($A_{d}$, $\alpha_{d}$ and $T_a^{d}$), and excitation ($A_{ex}$, $\alpha_{ex}$ and $T_a^{ex}$) are obtained from literature and are summarized in Table \ref{tab:plasmaparams}. $\mathrm{N_2 (ex)}$ here refers to a specie that lumps electronic and vibrationally excited states with a rate corresponding to the total excitation cross section as described in You et al. \cite{you2014simulation}. Other electron impact processes with greater granularity can be easily added to this model which is part of our future work, while we utilize a simplified model in this paper for brevity. We also note that these excited species tend to have relaxation times $\sim$ 4 ms at 1500 K (and longer at lower temperatures) according to experimental correlations \cite{castela2016modelling}, which is much larger than charge relaxation times ($\sim$ 5-10 $\mu$s) considered in this work. Therefore, appreciable relaxation of excited states to ground states may not happen during charge relaxation but can potentially happen between collisions, which is a subject of our future investigations.

Since the discharge under consideration will be at atmospheric pressure, ambipolar diffusion will be dominant \cite{lieberman1994principles}, for which the Eigen solution for the ion/electron density is of the form:
\begin{eqnarray}
    n(x)=n_0 cos(\beta (x-z/2))\,\,\,\forall\,\,\, x \in (0,z)\\
    \beta^2=\pi^2/z^2 = k_{iz}(T_e) N_G/D_a \label{eq:ambipolar} \\
    D_a=D_i \left(1+T_e/T_g\right) \label{eq:ambidcoeff}
\end{eqnarray}
where, $n_e=n_i=n_0$ is the bulk plasma density that equals the electron and ion densities ($n_e$ and $n_i$) under the quasineutral assumption, and $D_a$ is the ambipolar diffusion coefficient that depends on ion diffusion coefficient, electron and gas temperature, as given by Eq.\ \ref{eq:ambidcoeff}. Eq.\ \ref{eq:ambipolar} when rearranged becomes a non-linear equation for finding the electron temperature:
\begin{equation}
    k_{iz}(T_e) N_G z^2 - \pi^2 D_i(1+T_e/T_g)=0
    \label{eq:pb}
\end{equation}

We can formulate another equation to obtain electron/ion density ($n_e$) using energy balance. It should be noted that every point along the Paschen curve from $z_1$ to $z_t$ corresponds to a particle potential curve for a given charge as shown in Fig.\ \ref{(fig:paschen_img_pot}(a). The power deposited into the plasma when the particle moves along the Paschen curve is the time rate of change of the particle's self energy:
\begin{eqnarray}
    P_{pl}=\frac{d}{dt}\left(\frac{q(z)^2}{8 \pi \epsilon_0 r}\right) \label{eq:self_enrg}
\end{eqnarray}
Here we assume the particle's charge is spread on a spherical surface thus giving it a self energy of $E_{self}=\frac{q^2}{8 \pi \epsilon_0 r}$. Self energy is defined as the work done in charging the particle to a specified charge starting from zero charge, and corresponds to the energy of a spherical self-capacitor.
This power is deposited into the plasma in the form of both ion and electron Joule heating. A large portion of this power going into ion heating results in gas temperature rise \cite{sitaraman2011gas, sitaraman2012simulation, deconinck2009computational} which is typical of microdischarges with gap distances on the order of 10-100 $\mu$m. The electron Joule heating which is a fraction ($f_{EJ}$) of the total power can be assumed to be balanced by electron elastic and inelastic collision pathways as well as bounding surface electron energy losses:
\begin{align}
    f_{EJ} P_{pl}= \frac{3}{2} n_e k_B (T_e-T_g) \frac{2 m_e}{m_g} \nu_{eg} \Omega + \nonumber \\ 
     \left(k_{iz} N_G n_e E_{iz} +
    k_{d} N_G n_e E_{d} + k_{ex} N_G n_e E_{ex} \right) \Omega 
    +2 \Gamma_z (2 k_B T_e) A_{coll} \label{eq:eb1}\\
    \Gamma_z = D_a \beta n_e \label{eq:ambiflux}
\end{align}
Here, $\Omega=A_{coll} z$ is assumed to be the discharge volume, assuming a cylindrical discharge is formed with the collisional area as the cross section and distance from the surface as the height. \revise{The choice of $A_{coll}$ as the cross-sectional area for the discharge is based on evidence from recent studies that indicated localized spots and patches of charge formed during tribocharging \cite{preud2023tribocharging,lacks2019long}.}
The first term in Eq.\ \ref{eq:eb1} is the electron elastic collisional loss \cite{sitaraman2011gas} wherein $T_g$ is the gas temperature, $m_e$ is electron mass, $m_g$ the background gas mass (here for $\mathrm{N_2}$), and $\nu_{eg}$ is the electron-gas collision frequency. The second term is the electron inelastic loss \cite{sitaraman2011gas} through ionization, dissociation and excitation, and the last term is associated with energy loss at the bounding surfaces \cite{lieberman1994principles}.
The factor of 2 in the area loss term in Eq.\ \ref{eq:eb1} corresponds to losses on both boundaries (at both particle and wall) and an energy of $2 k_B T_e$ is lost per electron assuming a Maxwellian velocity distribution function for the electrons. $\Gamma_z$ is the electron flux obtained from ambipolar diffusion theory \cite{lieberman1994principles} as given by Eq.\ \ref{eq:ambiflux}.

The plasma power from Eq.\ \ref{eq:self_enrg}, can be further reduced to a function of 
distance $z$, assuming that every point the particle moves along the Paschen curve is on an 
intersection with an image charge potential equation (Eq.\ \ref{eq:imcharge}). This constraint is as 
shown in Eq.\ \ref{eq:ps_img}. We can 
also convert the time derivative to a distance derivative using a constant rebound velocity $v_f=e v_i$, where $e$ is the restitution coefficient and $v_i$ is the incident particle velocity:
\begin{align}
    P_{pl}=\frac{d}{dz}\left(\frac{q(z)^2}{8 \pi \epsilon_0 r}\right) v_f\\
          = \frac{v_f}{8 \pi \epsilon_0 r} \frac{d}{dz} (q(z)^2)\\
          = \frac{v_f}{8 \pi \epsilon_0 r} \frac{d}{dz} \left(B_p(z) 2 \pi \epsilon_0 r \left(\frac{r}{z}+2\right)\right)^2
   \label{eq:eb2}
\end{align}
Therefore, the overall energy balance equation is obtained by equating Eq.\ \ref{eq:eb1} and \ref{eq:eb2}, thus 
providing a way to quantify electron density as a function of $z$ as the particle moves from $z_1$ to $z_t$:
\begin{align}
    f_{EJ} \frac{v_f}{4} \frac{d}{dz} \left(B_p(z) \left(\frac{r}{z}+2\right)\right)^2 
    = 
    3 k_B n_e (T_e-T_g) \frac{m_e}{m_g} \nu_{eg} \Omega + \nonumber\\
    \left(k_{iz} E_{iz} 
    + k_{d} E_{d} + k_{ex} E_{ex} \right) \Omega  N_G n_e  
    + 4 D_a \beta n_e k_B T_e A_{coll} \label{eq:eb}
\end{align}
The electron Joule heating fraction ($f_{EJ}$) is assumed to be 6\% from the work of Deconinck et al. \cite{deconinck2009computational}  on atmospheric pressure microdischarges. This fraction need to be further refined via multi-dimensional simulations of the discharge, which is part of our future efforts. We continue to show the sensitivity of plasma species densities to this parameter in section \ref{sec:singlepart}.
Eq.\ \ref{eq:pb} and \ref{eq:eb} can be solved for electron temperature and electron density that can now be directly used
in rate equations for species, which can be integrated over the time the particle spends during charge relaxation:
\begin{eqnarray}
    \frac{d}{dt} (n_N \Omega) = v_f \frac{d}{dz} (n_N \Omega) = 2 k_{d} N_G n_e \Omega \label{eq:ode21}\\
    v_f \left(\Omega \frac{d n_N}{dz} + n_N \frac{d \Omega}{dz} \right)=2 k_{d} N_G n_e \Omega \label{eq:ode22} \\
    \frac{d n_N}{dz} = \frac{1}{v_f}\left(2 k_{d} N_G n_e - \frac{n_N}{z}\right) \label{eq:ode2}
\end{eqnarray}
It should be noted that the discharge volume $\Omega$ changes as the particle recedes away from the surface and therefore the time rate of change of concentration with changing volume is accounted through the second term on the right hand side in Eq.\ \ref{eq:ode2}.
Integrating Eqs.\ \ref{eq:ode2} from $z_1$ to $z_t$ gives the amount of dissociated species formed after charge relaxation. A similar ordinary differential equation can be formulated for $\mathrm{N_2(ex)}$ state number density, given by:
\begin{eqnarray}
\frac{d n_{N_2(ex)}}{dz} = \frac{1}{v_f}\left(k_{ex} N_G n_e - \frac{n_{N_2(ex)}}{z}\right) \label{eq:ode3}    
\end{eqnarray}
The initial condition for electrons, N, and $\mathrm{N_2(ex)}$ species density for Eqs.\ \ref{eq:ode2} and \ref{eq:ode3} were set to a minimal value of 1e12 $\mathrm{\#/m^3}$ that corresponds to representative background ionization levels in atmospheric conditions \cite{bagheri2018comparison}. The value of this minimum density was found to have no appreciable impact on our results in the range of 1e10 to 1e14 $\mathrm{\#/m^3}$. 
\subsection{Numerical implementation}
The plasma model given by Eqs.\ \ref{eq:pb}, \ref{eq:eb}, \ref{eq:ode2}, and \ref{eq:ode3}, 
form a set of differential algebraic equations which need to be solved alongside Eq.\ \ref{eq:ps_img} that determines particle charge during relaxation between $z_1$ to $z_t$. The solution to these equations were obtained from 
a python based solver that utilized SciPy's \cite{virtanen2020scipy} non-linear root finder, \textit{fsolve}. A good initial guess was required 
for root finding especially for determining $z_1$ from Eq.\ 
\ref{eq:ps_img} and for finding $z_t$ and $q_t$ from Eq.\ \ref{eq:tangent1} 
and \ref{eq:tangent2}, for which a graphical method was used prior to 
\textit{fsolve}, without which a converged solution was difficult to obtain. The ordinary differential equations were solved using a forward Euler scheme with solutions of electron temperature and density obtained from the non-linear solves of Eqs.\ \ref{eq:pb} and \ref{eq:eb}. The general workflow is as shown below in algorithm \ref{algo:solvealgo}: 
\begin{algorithm}
\caption{Coupled charge relaxation plasma solve algorithm.}
\label{algo:solvealgo}
\begin{algorithmic}[1]
    \State find $q_t$ and $z_t$ corresponding to tangential conditions (Eqs.\ \ref{eq:tangent1} and \ref{eq:tangent2})
    \State find $z_1$ corresponding to post contact charge given incoming charge
    \State Discretize domain from $z_1$ to $z_t$ with $i=0,N-1$ points
    \State Initialize electron and species densities of size $N-1$
    \State Initialize electron temperature array of size $N$
    \State Solve Eq.\ \ref{eq:pb} for electron temperature $T_e(i=0)$ at $z_1$
    \For{i = 0 to N-2}
    \State Solve for electron temperature at $T_e(i+1)$
    \State average $T_e$ between $i$ and $i+1$: $\bar{T}_{e}=0.5(Te(i)+Te(i+1))$
    \State find particle charge $q(i+1)$ at $z(i+1)$ from Eq.\ \ref{eq:ps_img}
    \State find self energy difference $\delta e_s$ using $q(i)$ and $q(i+1)$
    \State solve for electron density $n_e(i)$ using $\bar{T}_e$ and $\delta e_s$
    \State Forward Euler update of N radical density using $ne(i)$ and $\bar{T}_{e}$ using Eq.\ \ref{eq:ode2} 
    \State Forward Euler update $\mathrm{N_2(ex)}$ density using $n_e(i)$ and $\bar{T}_{e}$ using Eq.\ \ref{eq:ode3}
\EndFor
\end{algorithmic}
\end{algorithm}


%% file: results.tex
\subsection{Single particle collision}
\input{singlepart}
\subsection{Collection of particles: impact of space charge}
\input{multipart}
\subsection{Collection of particles: Granular motion}
\input{granular}

%% file: singlepart.tex
\label{sec:singlepart}
\begin{figure}
    \centering
    \includegraphics[width=\linewidth]{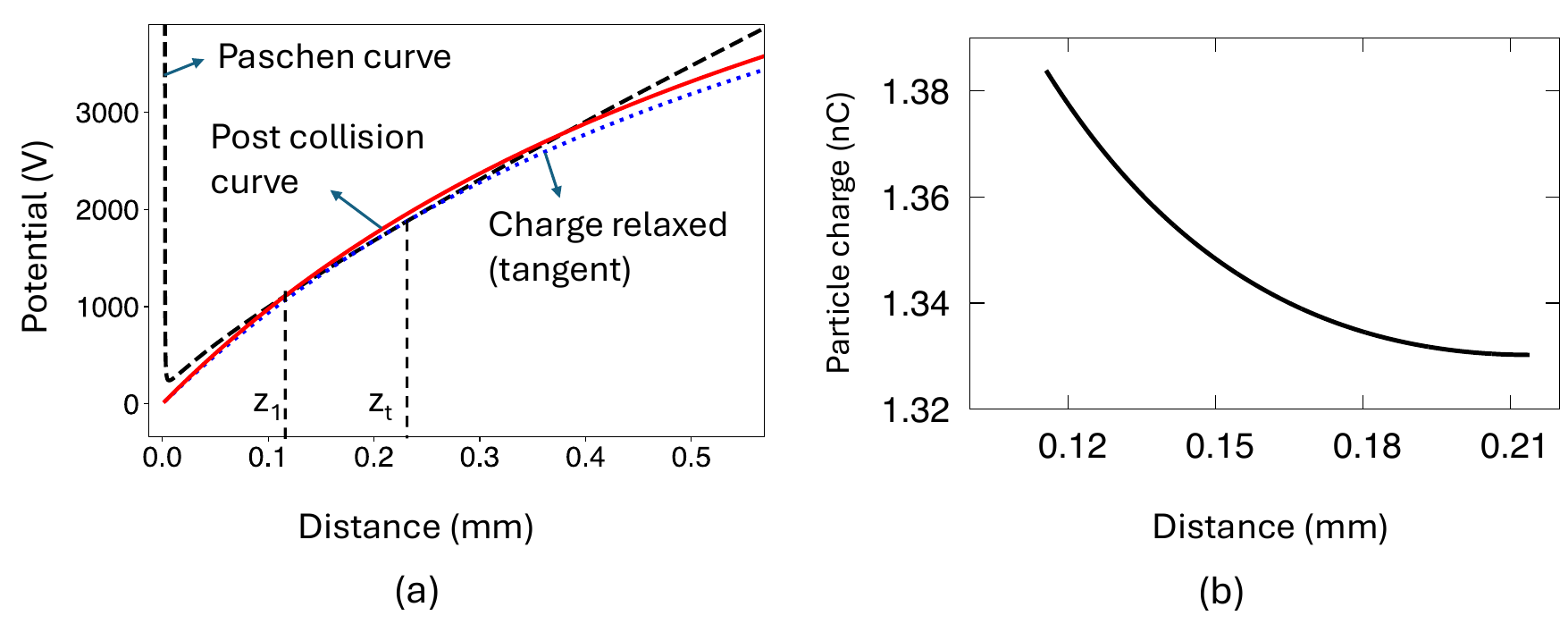}
    \caption{(a) Paschen curve for \ce{N2} along with the image potential curves for a 1.5 mm radius particle approaching the wall with charge $q_t=1.33\,\mathrm{nC}$ at 20 m/s along with the post collision potential curve at charge $q=1.38\,\mathrm{nC}$. Charge relaxation onset distances $z_1$ and tangential location $z_t$ are also indicated. (b) shows the variation of particle charge as it traverses the Paschen curve from $z_1$ to $z_t$.} 
    \label{fig:n2paschen}
\end{figure}
\begin{figure}
    \centering
    \includegraphics[width=0.5\linewidth]{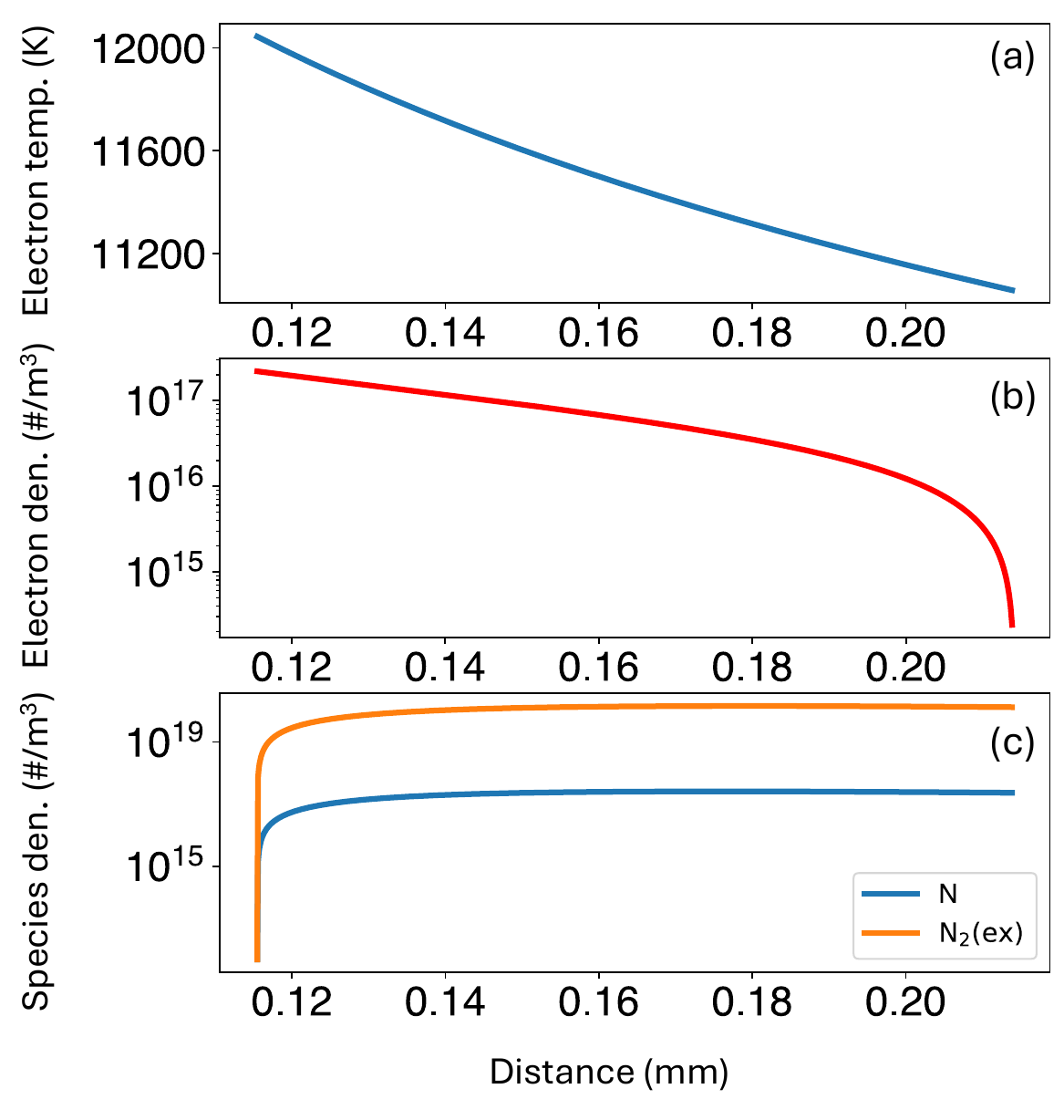}
    \caption{ 
    (a) electron temperature, (b) electron density and (c) N/$\mathrm{N_2 (ex)}$ density as the particle 
    traverses the Paschen curve from $z_1$ to $z_t$ in Fig.\ \ref{fig:n2paschen}(a).}
    \label{fig:n2params}
\end{figure}
As an example scenario, let us consider a particle repeatedly colliding with a surface and gaining charge. The particle would then at some point achieve a charge just enough so that its image potential curve is tangent to the Paschen curve as shown by the dotted line in Fig.\ \ref{fig:n2paschen}(a). A particle approaching the wall with this critical charge will then gain contact charge for which the post collision curve would intersect the Paschen curve as shown in Fig.\ \ref{fig:n2paschen}(a). Charge relaxation would happen and the particle would finally exit at location $z_t$ with charge $q_{t}$, obtained from Eqs.\ \ref{eq:tangent1} and \ref{eq:tangent2}. This process is repetitive if a particle is confined between two grounded conducting walls where the particle pre-collision charge will always remain at $q_t$. \revise{Pre charging of particles by external means will only expedite the initial process of achieving the critical tangential charge $q_t$, while the steady state repetitive process of charge relaxation and plasma discharge would remain the same.}The discharge events will cease when particle kinetic energy is dissipated from inelastic collisions with the wall as the contact charge which depends on collisional area, $A_{coll}$, tends to 0, \revise{unless external forcing such as in vibrating beds can maintain the particle's kinetic energy.}

\revise{We assume the physical parameters from Tables \ref{tab:tbparams} and \ref{tab:plasmaparams} for the cases studied in this work. 
We chose to study the collision of Teflon particles with aluminum surface, as this material combination has been studied extensively in tribocharging literature \cite{zhang2020material}. Teflon has a high electron affinity and is in the lower end of triboelectric series \cite{sotthewes2022triboelectric}  while \ce{Al} is a good electron donor and is at the higher end of the triboelectric series, resulting in a large effective work function difference.} \revise{The particle radius in this case was set to 1.5 mm based on range of sizes used in previous works on plasma catalysis \cite{chen2021plasma, bogaerts2019burning} and ongoing experiments in our lab.}
$q_{t}$, $z_1$ and $z_t$ were calculated to be 1.33 nC, 115 $\mu$m, and 214 $\mu$m, respectively, for this system.
The charge gained during contact in this case was found to be approximately 54 pC from Eq.\ \ref{eq:tribocharge}. The particle charge history when it traverses the Paschen curve between $z_1$ and $z_t$ is shown in Fig.\ \ref{fig:n2paschen}(b), assuming that every point on the Paschen curve corresponds to an image potential curve of specified charge. As expected, the charge reduces and finally asymptotes to $q_t$, as the particle moves from $z_1$ to $z_t$.  
\begin{table}
    \centering
    \begin{tabular}{c|c|c}
     \hline
     \hline
     Parameter & Value & Reference \\
     \hline
     \hline
     Critical gap $\delta_c$ & 88e-9 & \cite{tan2023valuations} \\
     Tribocharging efficiency $k_c$      & 1.0 & \cite{tan2023valuations} \\
    Particle Young's modulus (Teflon) & 5.35e8 GPa & \cite{armitage2022investigating} \\
    Particle Poisson's ratio (Teflon) & 0.25 & \cite{armitage2022investigating} \\
    Particle density (Teflon) & 2200.0 $\mathrm{kg/m^3}$ & \cite{nist_teflon}\\
    Work function (Teflon) & 5.8 eV & \cite{armitage2022investigating} \\
    Wall Young's modulus (Al) & 68.9e9 GPa & \cite{armitage2022investigating} \\
    Wall Poisson's ratio (Al) & 0.275 & \cite{armitage2022investigating} \\
    Wall density (Al) & 2700.0 $\mathrm{kg/m^3}$ & \cite{nist_al}\\
    Work function (Al) & 4.26 eV & \cite{armitage2022investigating} \\
    particle radius & 1.5 mm & \\
    Initial velocity & 20 m/s & \\
    Restitution coefficient & 0.9 & \\
    \hline
    \end{tabular}
    \caption{Tribocharging and material parameters used in this work.}
    \label{tab:tbparams}
\end{table}
\begin{table}
    \centering
    \begin{tabular}{c|c|c}
     \hline
     \hline
     Parameter & Value & Reference \\
     \hline
     \hline
    Pressure & 101325.0 Pa &  \\
    Temperature & 300 K & \\
    Paschen curve $V_{min}$ for \ce{N2} & 240.0 V & \cite{raizer1997gas} \\
    Paschen curve $pd_{min}$ for \ce{N2} & 4.9 mmTorr &  \cite{raizer1997gas} \\
    $E_{iz}$ & 15.6 eV & \cite{yuan2003computational}\\
    $E_{d}$ & 9.757 eV & \cite{yuan2003computational}\\
    $E_{ex}$ & 6.17 eV & \cite{you2014simulation} \\
    $\alpha_{iz}$ & -0.3 & \cite{yuan2003computational}\\
    $\alpha_{d}$ & -0.7 & \cite{yuan2003computational}\\
    $\alpha_{ex}$ & 0.0 & \cite{you2014simulation} \\
    $A_{iz}$ & 4.483e-13 $\mathrm{m^3}$/\#/s & \cite{yuan2003computational}\\
    $A_d$ & 1.959e-13 $\mathrm{m^3}$/\#/s & \cite{yuan2003computational}\\
    $A_{ex}$ & 4.05e-15 $\mathrm{m^3}$/\#/s & \cite{you2014simulation}\\ 
    $T_a^{iz}$ & 1.81e5 K & \cite{yuan2003computational}\\
    $T_a^d$ & 1.132e5 K & \cite{yuan2003computational}\\
    $T_a^{ex}$ & 6.2e4 K & \cite{you2014simulation} \\
    $f_{EJ}$ & 0.06 & \cite{deconinck2009computational}\\
    $D_i$ & 6.388e-6 $\mathrm{m^2/s}$ & \cite{viehland1995transport} \\
    $\nu_{eg}$ & 482 to 5870 GHz & from BOLSIG+ \cite{hagelaar2005solving}\\
    \hline
    \end{tabular}
    \caption{Plasma and gas phase parameters used in this work.}
    \label{tab:plasmaparams}
\end{table}

Fig.\ \ref{fig:n2params} shows the plasma parameters during dielectric breakdown as the particle 
traverses between $z_1$ and $z_t$ in Fig.\ \ref{fig:n2paschen}(a). 
As the charge relaxes along the Paschen curve, the electron temperature and electron density also 
follow a similar trend as Fig.\ \ref{fig:n2paschen}(b). Less energy is deposited in the plasma as the charge reduces to the final tangential charge, $q_t$. Electron temperatures greater than 1 eV ($\sim$ 11604 K) is seen in Fig.\ \ref{fig:n2params}(a) that is similar to average temperatures observed in recently studied atmospheric pressure non-equilibrium plasma catalytic systems \cite{hong2017kinetic,van2020plasma}.
Fig.\ \ref{fig:n2params}(c) shows the N and $\mathrm{N_2 (ex)}$ number densities respectively, indicating a rapid increase initially from the high electron densities at the onset of charge relaxation. Appreciable peak electron, radical and excited species concentrations $\sim$ 1E17, $\sim$ 2E17 $\mathrm{\#/m^3}$ and $\sim$ 1.5E20  $\mathrm{\#/m^3}$, respectively are observed at the end of charge relaxation, respectively. Fig.\ \ref{fig:n2params}(c) also indicate three orders of magnitude higher production of excited species compared to dissociated \ce{N2} which is typical of discharges operating in the 1-3 eV range \cite{rouwenhorst2019vibrationally}. Enhanced production of excited states like in this case is considered as a more energy efficient pathway for plasma catalytic ammonia synthesis as opposed to increasing \ce{N2} dissociation \cite{lin2024kinetic}. 
\begin{figure}
    \centering
    \includegraphics[width=0.5\linewidth]{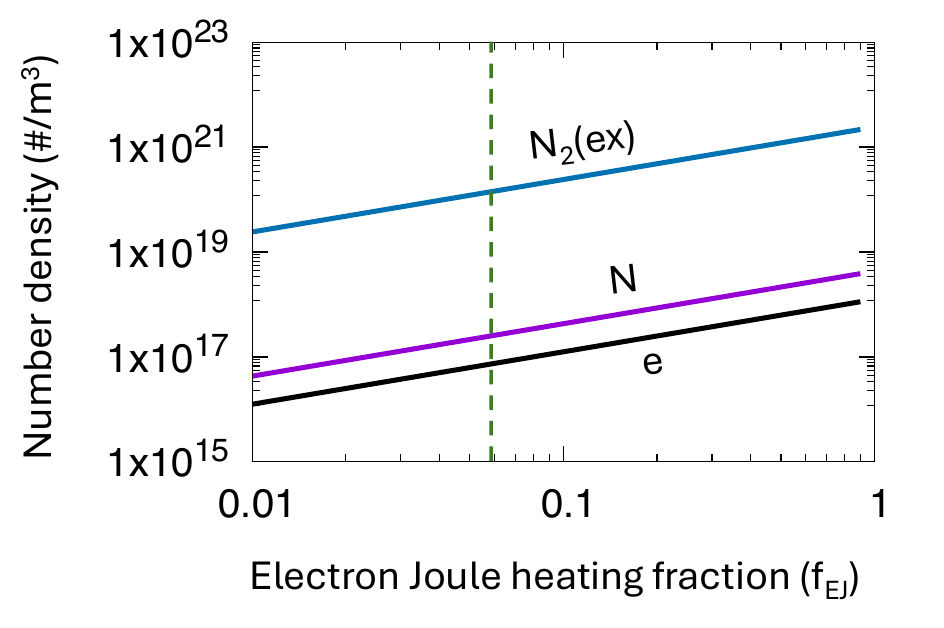}
    \caption{ 
    Sensitivity of average electron density (between $z_1$ and $z_t$), final N and $\mathrm{N_2(ex)}$ densities (at $z_t$) to electron Joule heating fraction $f_{EJ}$ from Eq.\ \ref{eq:eb}. The dashed line indicates the baseline value of 6\% chosen in this work based on microdischarge simulations \cite{deconinck2009computational}.}
    \label{fig:ejsens}
\end{figure}

Fig.\ \ref{fig:ejsens} shows the sensitivity of electron, N and $\mathrm{N_2(ex)}$ densities to the choice of electron Joule heating fraction $f_{EJ}$ from Eq.\ \ref{eq:eb}. The plasma species densities are directly proportional to $f_{EJ}$ which essentially determines the fraction of electrical energy from charge relaxation going towards increasing electron energy. The peak species densities from charge relaxation here are comparable to densities observed in plasma catalytic dielectric barrier discharge (DBD) reactors for ammonia synthesis. For example, Van't Veer et al. \cite{van2020plasma} report electron, N, and vibrational \ce{N2} densities in the range 1E16-1E18, 1E17-1E20, and 1E21-1E23 $\mathrm{\#/m^3}$ respectively, in a single DBD microdischarge event. Furthermore, we can surmise that multiple triboplasmas from many particles will enable the formation of larger quantities of activated species, as will be further analyzed in section \ref{sec:granular}. 

It is clear from the above illustrative example of a single particle collision using realistic material and plasma parameters that triboplasmas generated during charge relaxation can indeed serve as a pathway for the production of excited \ce{N2} species that can enable catalytic conversion processes. In the upcoming sections, we continue to answer the question about the impact of multiple particles, sensitivity to particle and material parameters, and particle velocity distributions.

%% file: multipart.tex
\label{sec:collectpart1}
\begin{figure}
    \centering
    \includegraphics[width=0.8\linewidth]{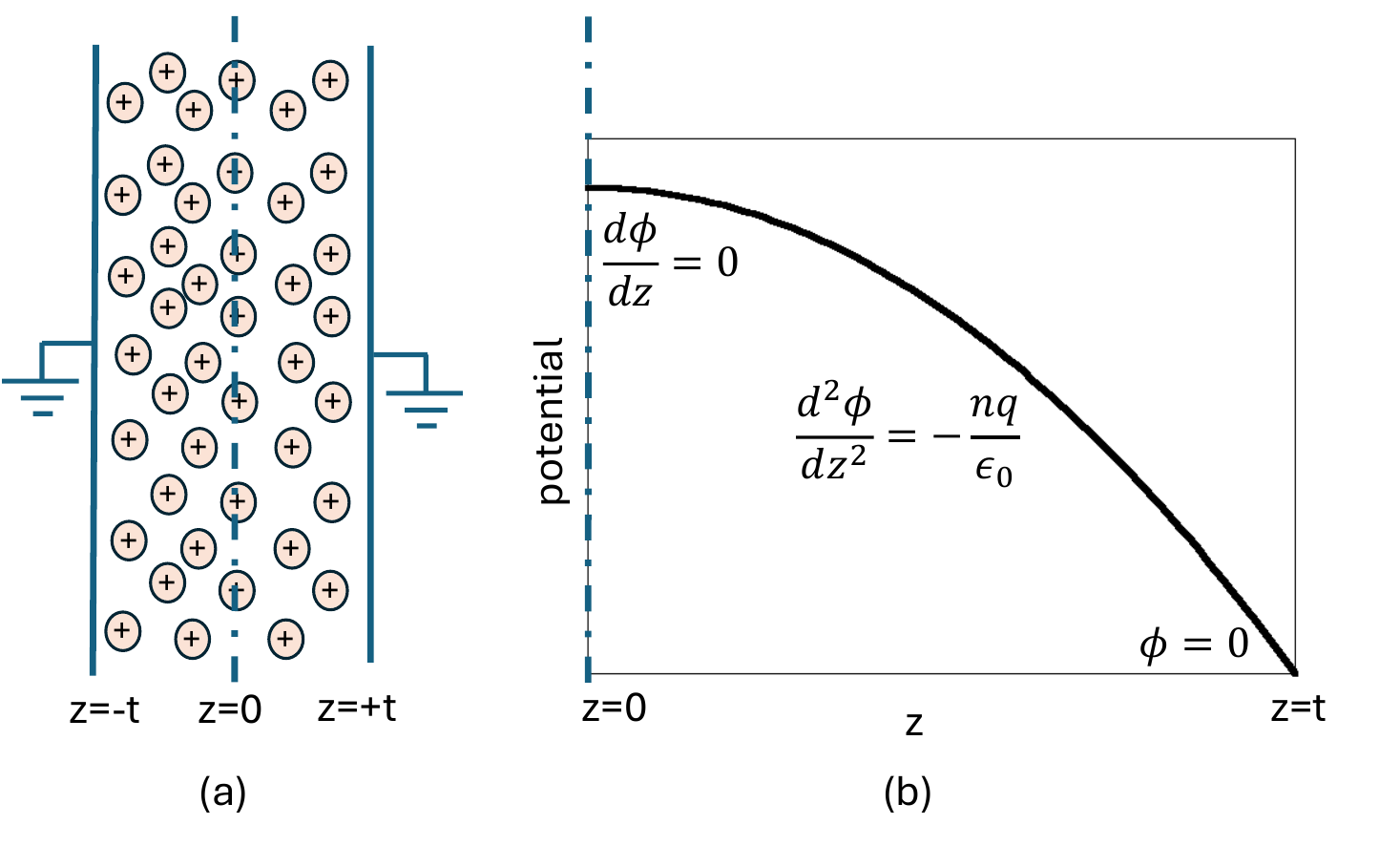}
    \caption{(a) schematic of particles confined between two grounded infinite parallel plates and (b) potential solution with constant space charge density satisfying boundary conditions as shown here from Gauss's law.}
    \label{fig:spacecharge}
\end{figure}
In this section we derive the coupled charge-relaxation-plasma model when a collection of particles are charged from impact on the walls, as opposed to the single particle case studied in the previous section.

Consider a system where particles are confined between two parallel walls as they are repeatedly charged by contact, as shown in Fig.\ \ref{fig:spacecharge}(a). We assume for simplicity that all particles possess the same directed velocity towards the walls and collisions among particles are neglected. The inclusion of velocity distribution functions is studied in the upcoming section (section \ref{sec:granular}). 
As the space charge in the bulk of the domain builds up with multiple particles, a potential field is established within the gap. This potential field can be obtained from Gauss's law with a symmetry boundary condition in the middle of the gap while grounded (potential equal to zero) boundary condition at the walls as shown in Fig.\ \ref{fig:spacecharge}(b):
\begin{eqnarray}
    \frac{d^2\phi}{dz^2}=-\frac{nq}{\epsilon_0} \in (0,t)\\
    \frac{d\phi}{dz} \bigg\rvert_{z=0}=0\,\,\,\,\,\phi|_{z=t}=0
\end{eqnarray}
Here $n$ is the number density of particles and $q$ is the charge possessed by each particle. A parabolic solution for $\phi$ or the bulk potential $V_b(z)$ from Eq.\ \ref{eq:tribocharge} can be easily obtained assuming a spatially independent number density and charge, given by: 
\begin{eqnarray}
    \phi(z)=V_b(z)=\frac{nq}{2\epsilon_0}(t^2-z^2) \label{eq:channelpot}
\end{eqnarray}
This potential will change sign if the particles are charged negatively, but the magnitude of potential difference between the wall and particles would follow the curve as shown in Fig.\ \ref{fig:spacecharge}(b).
The condenser model equation for contact charge gained during collision that includes bulk potential effects is a modification to Eq.\ \ref{eq:tribocharge} as shown below:
\begin{eqnarray}
    \delta q=k_c \frac{\epsilon_0 A_{coll}}{\delta_c} \left(V_c - V_{img}(q,\delta_c) - V_b(q,\delta_c)\right) \label{eq:tribochargebulk}
\end{eqnarray}

The space charge electric field is always directed away from charge transfer, no matter the sign of the charge gained by the particle, thus reducing the contact charge gained by the particle. On the other hand, the space charge field results in higher voltages compared to just the image potential and therefore increases the chance of intersections with the Paschen curve. 

Equations \ref{eq:tangent1} and \ref{eq:tangent2} for obtaining $q_{t}$ and $z_t$, the particle charge and location for tangential condition with the Paschen curve $B_p(z)$, need to be modified with the space charge potential $V_b(z)$, given by:
\begin{eqnarray}
    \frac{dV_{img}}{dz}(q_{t},z_t) + \frac{dV_b}{dz}(q_{t},z_t)=
    \frac{d B_{p}}{dz}(z_t) \label{eq:tangent3}\\
    V_{img}(q_{t},z_t)+V_b(q_{t},z_t)=B_p(z_t) \label{eq:tangent4}
\end{eqnarray}
\begin{figure}
    \centering
    \includegraphics[width=0.5\linewidth]{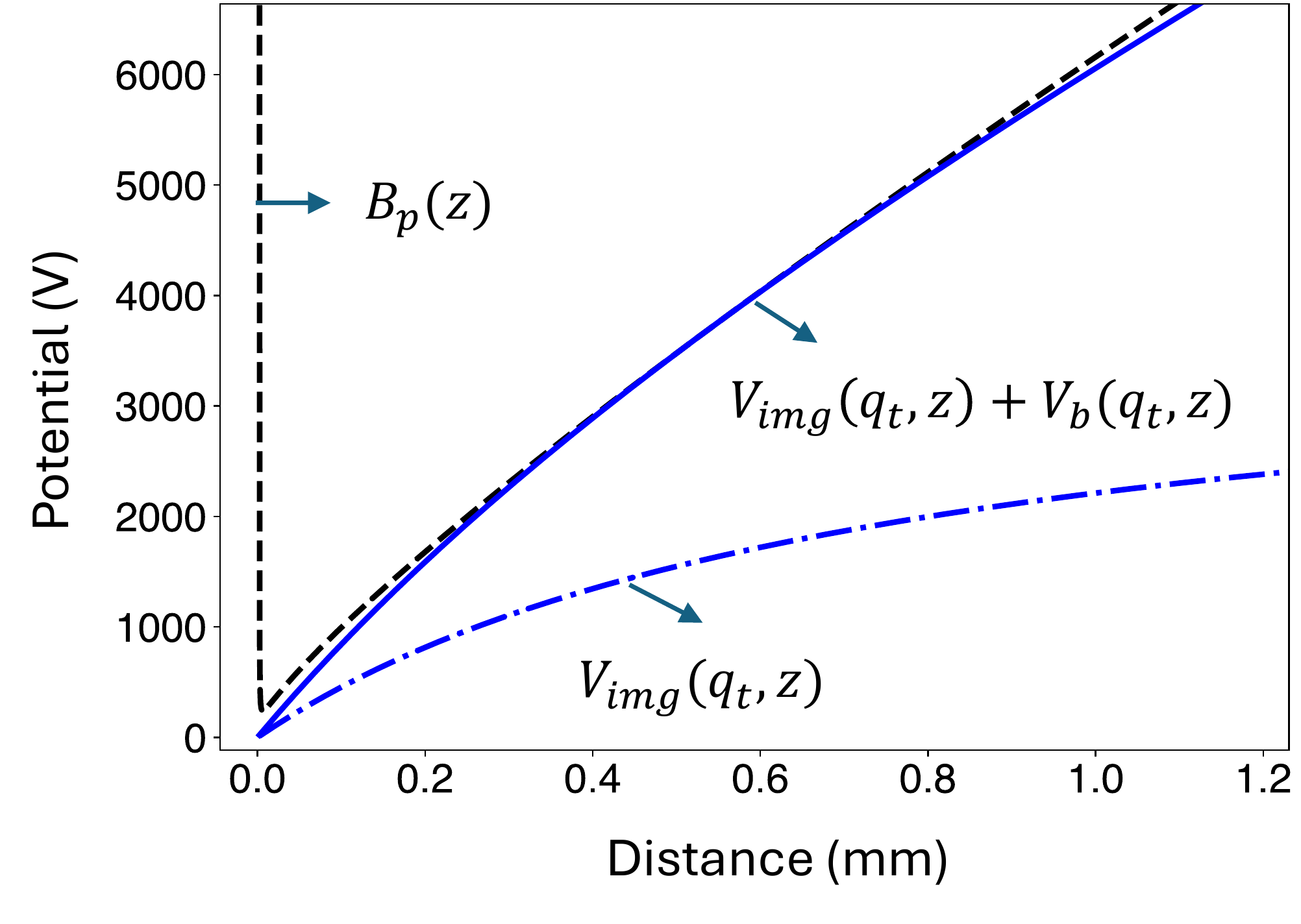}
    \caption{Comparison of total potential (image and bulk) versus image potential alone when tangent condition achieved from Eqs. \ref{eq:tangent3} and \ref{eq:tangent4}.}
    \label{fig:imgbulkpaschen}
\end{figure}
\begin{table}
    \centering
    \begin{tabular}{c|c}
     \hline
     \hline
     Parameter & Value \\
     \hline
     \hline
     Solids fraction ($\epsilon$) & 0.01 \\
     Particle number density $\left(n = \frac{\epsilon}{4/3 \pi r^3}\right)$ & 7.07e5 $\mathrm{\#/m^3}$ \\
     average distance between particles $\left(\lambda=n^{-1/3}\right)$ & $7.5 r = 11.25\,\mathrm{mm}$ \\ 
     Channel width ($2t$) & 0.015 m \\
    \hline
    \end{tabular}
    \caption{Parameters for coupled charge relaxation-plasma model with multiple particles.}
    \label{tab:multipartparams}
\end{table}
As an example baseline case we assume a solids fraction of 1\% and channel width that is 100 times the radius of the particle ($r=1.5\,\mathrm{mm},\,\, 2t=0.15\,\mathrm{m}$). The particle number density and average distance between particles calculated based on the solids fraction is as shown in Table \ref{tab:multipartparams}. The solids fraction is assumed to be a small value so that the collisionless assumption holds reasonably true with an average distance between particles that is 7 times the particle radius in this case. We assume that the particles first repetitively collide with the wall until their charge is large enough to yield a total potential that touches the Paschen curve. Plasma discharges happen in subsequent collision after this critical charge is achieved. With the parameters from Table \ref{tab:multipartparams}, the tangential charge and location to the Paschen curve ($q_t$ and $z_t$) are calculated using Eqs.\ \ref{eq:tangent3} and \ref{eq:tangent4}, for which the total particle potential curve is shown in Fig.\ \ref{fig:imgbulkpaschen}. The solution to $q_{t}$ and $z_t$ were calculated to be 0.65 nC and 534 $\mu$m, respectively. From Fig.\ \ref{fig:imgbulkpaschen}, it is seen that there is a significant contribution from the bulk space charge which results in a lower particle charge for achieving the tangential condition with the Paschen curve. The single particle scenario yielded $q_t=1.3\,\mathrm{nC}$ which is about twice as much compared to this multi particle case.
\begin{figure}
    \centering
    \includegraphics[width=0.7\linewidth]{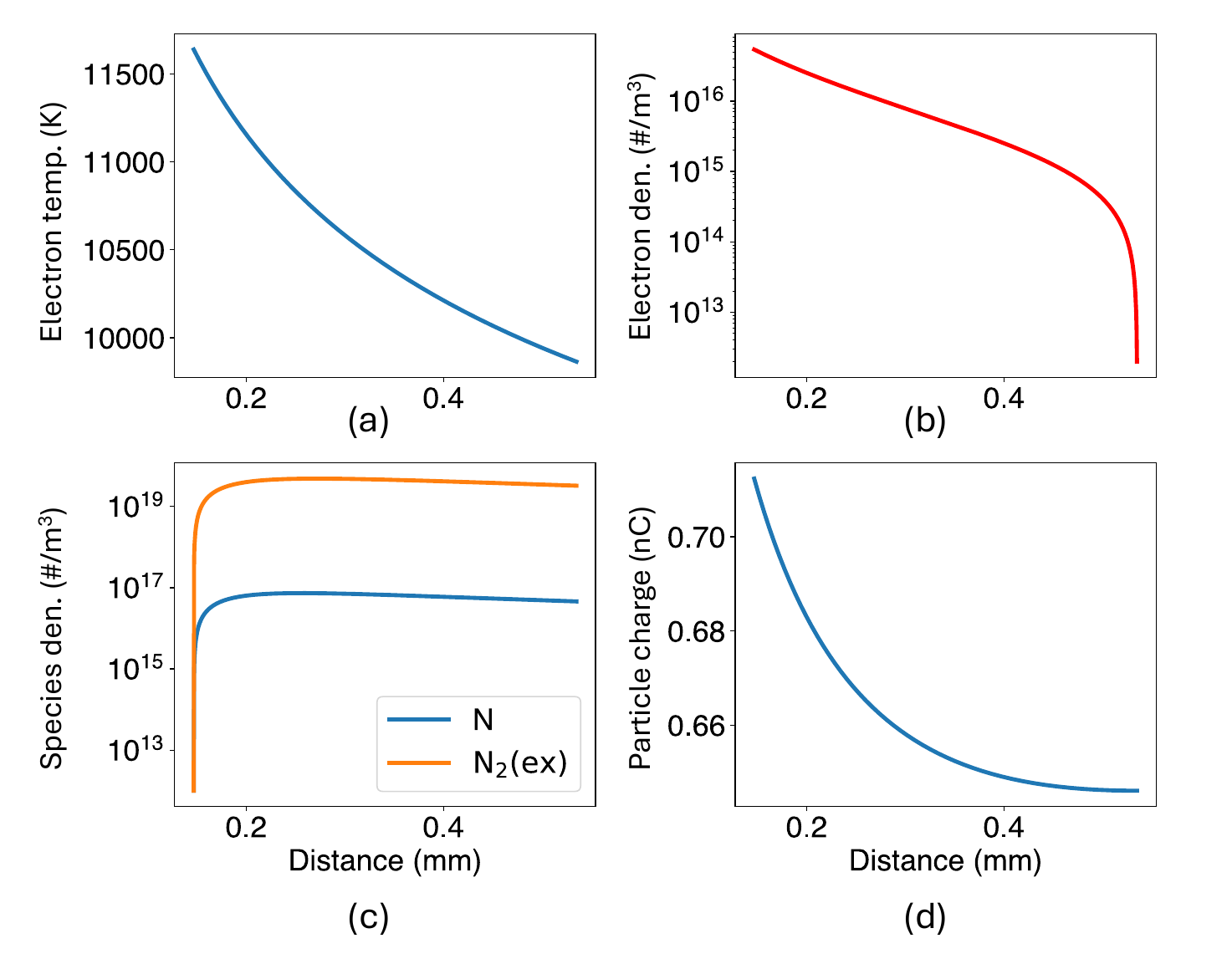}
    \caption{(a) electron temperature, (b) electron density, (c) species densities (N and $\mathrm{N_2(ex)}$), and (d) particle charge as a function of distance traversed along the Paschen curve when space charge effects are included.}
    \label{fig:multip_plasmaparams}
\end{figure}
We continue to use the same plasma model per particle as in the single particle case for this multi particle scenario, with the added potential associated with the space charge. Fig. \ref{fig:multip_plasmaparams} shows the electron temperature, electron number density, N/$\mathrm{N_2(ex)}$ number density and particle charge as a function of distance as the particle traverses the Paschen curve from $z_1$ to $z_t$. The trends in these parameters are similar to single particle case (see Fig.\ \ref{fig:n2params}), with the main difference being the lower post collision charge of $\sim$ 0.7 nC (Fig.\ \ref{fig:multip_plasmaparams}(d)) in this case compared to $\sim$ 1.4 nC in the single particle case. Lower electron, N radical and $\mathrm{N_2 (ex)}$ densities are achieved in this multiparticle case compared to the single particle case as shown in the averaged and peak values quantified in Table \ref{tab:compareplasmaparams}. 
The average electron temperature is more or less the same between single and multiple particle cases mainly determined by ambipolar diffusion. The average electron density on the other hand is approximately seven times lower with the inclusion of space charge effects while N and $\mathrm{N_2(ex)}$ number densities are also reduced by factor of three. This diminishing effect on plasma parameters in the presence of space charge potential is mainly because of the reduced initial pre-collision charge ($q_t$) compared to the single particle case resulting in lower energy dissipated into the discharge.
\begin{table}
    \centering
    \begin{tabular}{c|c|c}
     \hline
     \hline
     Parameter & Single particle & multiple particles \\
     \hline
     \hline
     Average electron temperature (K) & 11485  & 10516 \\
     Average electron density ($\mathrm{\#/m^3}$) & 7.5e16 & 1.05e16 \\
     Final N density (($\mathrm{\#/m^3}$) &  2.6e17 & 7.33e16 \\
     Final $\mathrm{N_2(ex)}$ density ($\mathrm{\#/m^3}$) & 1.45e20 & 4.84e19 \\
    \hline
    \end{tabular}
    \caption{Comparison of plasma parameters (electron temperature, electron, N, $\mathrm{N_2(ex)}$ densities) between single and multiple particle models.}
    \label{tab:compareplasmaparams}
\end{table}

This section demonstrated the importance of space charge effects with multiple particles and showed that bulk potentials reduced the tangential particle charge ($q_t$) compared to an isolated particle case from section \ref{sec:singlepart}. This reduced particle charge has a diminishing effect on the production of plasma species, but multiple particles also increase overall discharge volume compared to a single isolated particle. These competing effects are considered in an upcoming section on granular systems with velocity distributions (section \ref{sec:granular}), while we move on to understanding the sensitivity of overall plasma species densities to particle parameters (e.g., radius, velocity) and reactor scale parameters (e.g., channel width and solids fraction) in the upcoming section.
\subsection{Sensitivity studies}
\begin{figure}
    \centering
    \includegraphics[width=0.7\linewidth]{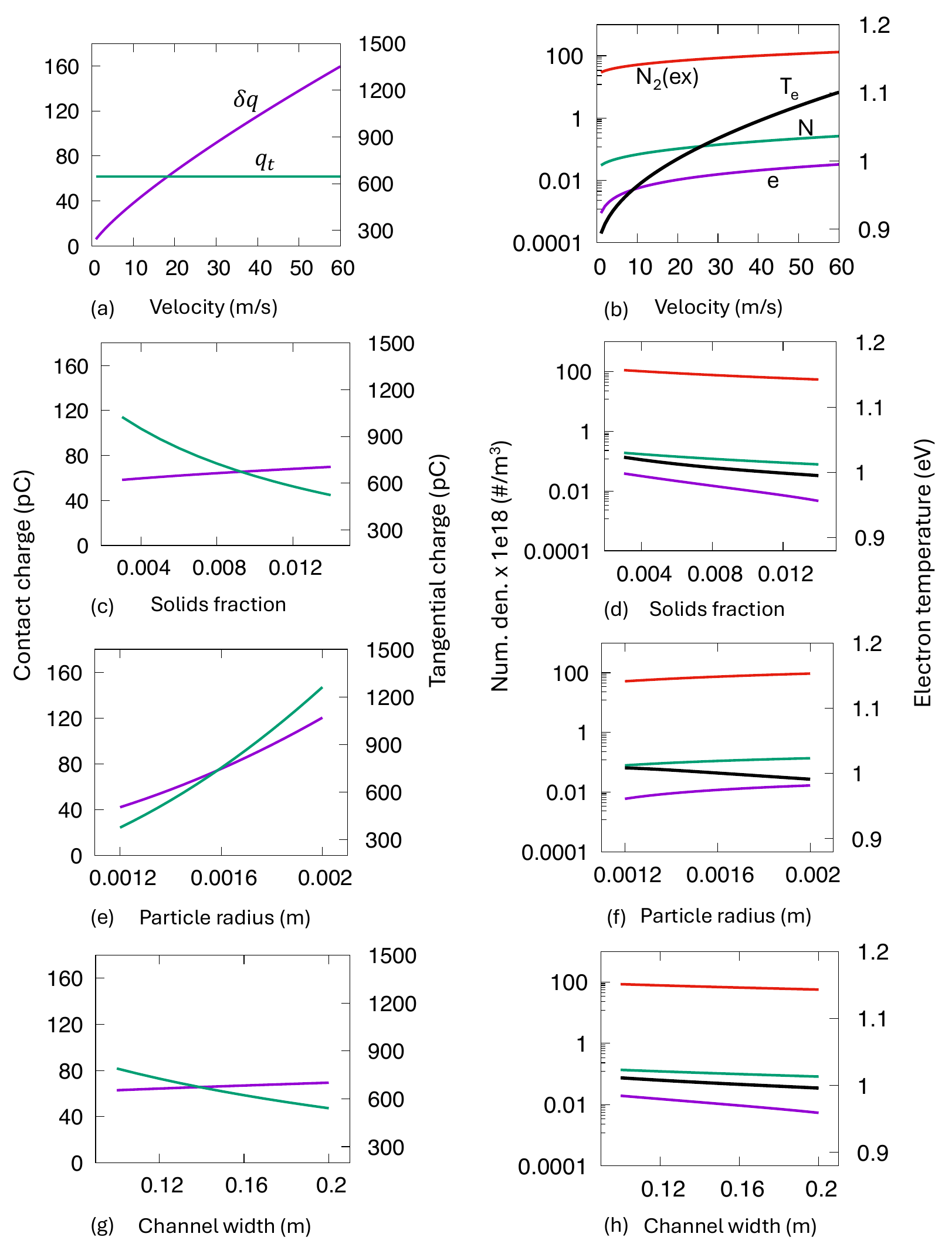}
    \caption{Sensitivity of contact charge ($\delta q$) shown on left axis and tangential charge $q_t$ shown on right axis with (a) particle velocity, (c) solids fraction, (d) particle radius and (g) channel width. (b), (d), (f) and (h) show variation of plasma parameters (electron, $\mathrm{N_2(ex)}$ and N density on left axis and electron temperature on right axis) with particle velocity, solids fraction, particle radius, and channel width, respectively.}
    \label{fig:multip_sens}
\end{figure}
In this section, we perform sensitivity studies for tribocharge and plasma parameters to four important parameters, that include particle velocity, solids volume fraction, particle radius and channel width using the more complete multiple particle model from the previous section. 

Figures \ref{fig:multip_sens}(a)-(h) shows these sensitivity studies where one parameter is varied while keeping others constant. The base parameters for velocity, solids fraction, particle radius, and channel width are 20 m/s, 0.01, 1.5 mm, and 0.15 m, respectively, as shown in Tables \ref{tab:tbparams} and \ref{tab:multipartparams}. 

Figures \ref{fig:multip_sens}(a) and (b) indicate a direct variation of contact charge and plasma parameters with velocity. Higher velocity results in greater compression of the particle during contact and increases the contact area ($A_{coll}$), resulting in more contact charge being transferred. Higher charge therefore results in higher number densities of plasma species (e, N, $\mathrm{N_2(ex)}$). Electron temperature remains close to 1 eV with minor variations resulting from minor reductions in charge relaxation onset distance ($z_1$) that reduces with higher contact charge, as can be seen from Fig.\ \ref{(fig:paschen_img_pot}. The tangential charge ($q_t$) is unaffected with particle velocity as it depends only on bulk and image potential that are independent of velocity.

Figures \ref{fig:multip_sens}(c) and (d) show the impact of solids fraction on particle charge and discharge parameters. Higher solids fraction results in higher space charge voltages which reduces the tangential charge ($q_t$). Lower tangential charge has a lower impact on contact potential difference thus resulting in minor increasing variations in contact charge as shown in Fig.\ \ref{fig:multip_sens}(c). However, the reduction in $q_t$ is the main reason behind an inverse trend in plasma species densities and electron temperature, mainly arising from the increased impact of space charge, as discussed in the previous section. 

Figures \ref{fig:multip_sens}(e) and (f) show a direct dependence of contact and tangential charge on particle radius. Larger particle radius have two reinforcing effects on contact charge: increased contact surface area ($A_{coll}$) and reduced image charge voltage at contact from inverse-square dependence of particle radius. The tangential charge also increases with particle radius because a larger charge is required to meet the Paschen curve due to the reduction in image voltage from inverse-square dependence on radius. There is a weak dependence of plasma parameters on particle radius with slight increases at larger particle radii from the increased contact charge and longer relaxation distance from the higher tangential charge. 

Figures \ref{fig:multip_sens}(g) and (h) show the sensitivity of particle charging and plasma parameters on channel width ($2t$). Larger channel width increases the space charge voltage contribution (increases as square of $t$) thus reducing the tangential charge required for meeting the Paschen curve as indicated in Fig.\ \ref{fig:multip_sens}(g). There is a weak increase in contact charge mainly because of the reduced contribution from image charge potential at the critical contact distance ($\delta_c$) due to lower $q_t$, even though there is an increase in space charge based potential. The plasma species density and electron temperature show an inverse dependence with increasing channel width mainly governed by the reduction in tangential charge $q_t$, that determines overall energy deposition into the plasma.

\begin{figure}
    \centering
    \begin{subfigure}{0.49\textwidth}
    \includegraphics[width=0.99\textwidth]{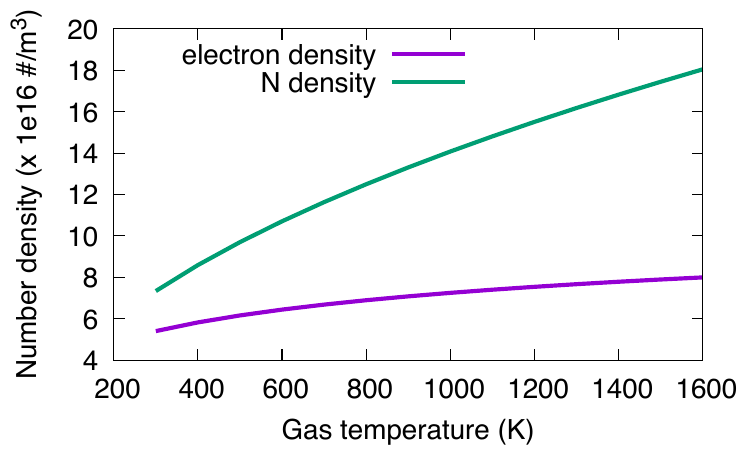}
    \caption{}
    \end{subfigure}
    \begin{subfigure}{0.49\textwidth}
    \includegraphics[width=0.99\textwidth]{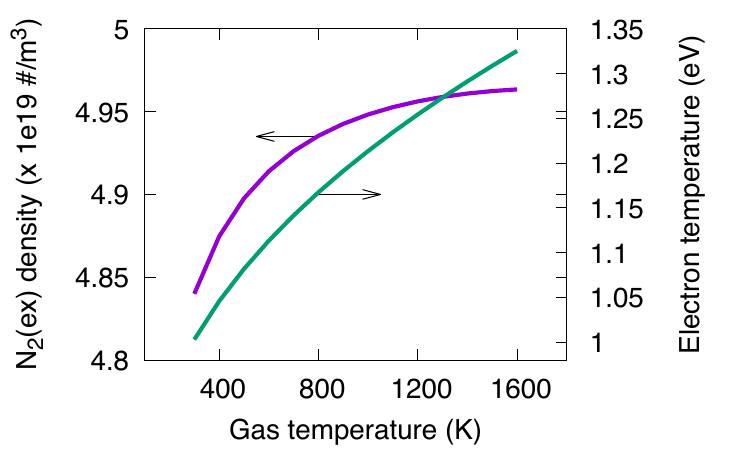}
    \caption{}
    \end{subfigure}
    \caption{Variation of (a) electron and N density, and (b) \ce{N2}(ex) density and electron temperature, as a function of gas temperature.}
    \label{fig:gastempsens}
\end{figure}

\revise{There are several mechanisms by which the gas temperature can be altered during tribocharging and subsequent plasma discharge, such as mechanical energy to heat dissipation during contact and microplasma assisted gas heating. In order to assess the impact of heating, we perform a sensitivity study of plasma parameters with varying gas temperature while keeping identical baseline parameters as the earlier cases. Figure \ref{fig:gastempsens}(a) shows variation of electron and N radical density with gas temperature while Fig.\ \ref{fig:gastempsens}(b) shows \ce{N2}(ex) density and electron temperature as a function of gas temperature. Since the pressure is kept fixed in these cases, higher gas temperature results in lower background gas densities. A lower gas density results in higher reduced electric fields thus increasing electron temperature and electron impact reaction rates. N density is observed to increase rapidly with higher gas temperature ($\sim$ 3X increase at 1600 K compared to baseline 300 K) while electron density and \ce{N2}(ex) density show modest increments, mainly due to the differences in reaction rates.}

Overall, it is seen that larger particles and higher particle velocities increase the production of active plasma species (N and $\mathrm{N_2(ex)}$) mainly from increased contact charging. On the other hand, higher solids fraction and channel width tend to reduce the densities of plasma species, mainly due to the effect of space charge fields. 
\revise{Increase in gas temperature while maintaining the same pressure is observed to favor higher plasma density and electron temperature.}
It should be noted that these sensitivities are on a per particle basis, which we later expand for a realistic granular system and predict overall plasma species production, in section \ref{sec:granular}.
\subsection{Plasma feasibility with varying gases}

Different gases and gas mixtures exhibit varying Paschen coefficients and it is important to assess whether a contact charge transfer would result in an intersection with the Paschen curve and subsequent plasma formation. In order to evaluate this feasibility, we check if the equilibrium incoming charge $q_{eqbm}$ obtained from the balance of contact potential difference $V_c$ and the sum of image ($V_{img}$) and bulk ($V_b$) potentials at critical gap distance exceeds the tangential charge $q_t$. $q_{eqbm}$ can be obtained from equating the amount of contact charge gained to 0 in Eq.\ \ref{eq:tribochargebulk} resulting in a linear equation for $q_{eqbm}$ given by:
\begin{eqnarray}
 V_{img}(q_{eqbm},\delta_c) + V_b(q_{eqbm},\delta_c)=V_c
 \label{eq:qeqbm}
\end{eqnarray}

As mentioned earlier, the particle has to charge to atleast $q_t$ so that the next collision would enable an intersection with the Paschen curve resulting in a plasma and charge relaxation back to $q_t$. Therefore, no plasma can be obtained if $q_{t}$ is greater than $q_{eqbm}$. This condition ($q_t = q_{eqbm}$) can now give bounds on the choice of particle/wall material for a given gas, so that a plasma based relaxation is possible. 
\begin{figure}
    \centering
    \includegraphics[width=0.99\linewidth]{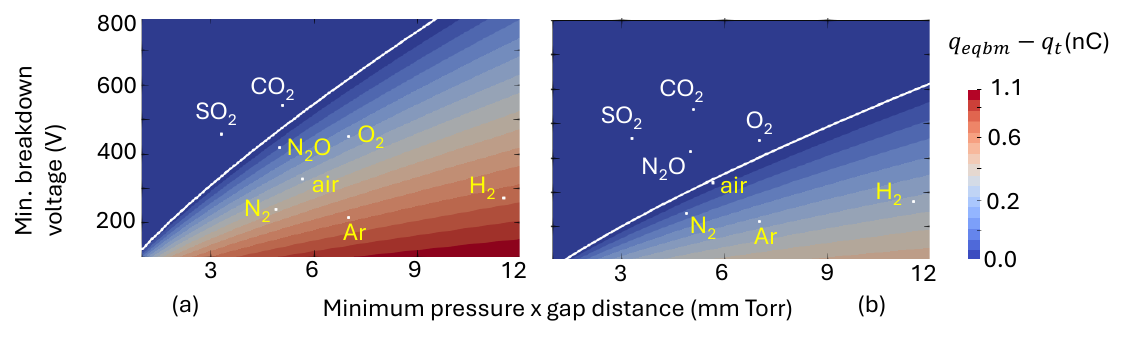}
    \caption{A two dimensional contour plot of minimum breakdown voltage ($V_{min}$) versus minimum pressure times discharge-gap ($pd_{min}$) colored by the difference between equilibrium charge ($q_{eqbm}$) from Eq.\ \ref{eq:qeqbm}) and tangential charge ($q_t$) along with a line that corresponds to $q_{eqbm}=q_t$ separating plasma feasible region from infeasible. The Paschen breakdown parameters for each gas is shown as a point for gases \ce{N2} \cite{raizer1997gas}, \ce{CO2} \cite{stumbo2013paschen}, \ce{N2O}, \ce{O2}, air, \ce{SO2}, \ce{Ar}, and \ce{H2} \cite{naidu1995}.}
    \label{fig:regimemap}
\end{figure}

This regime map is shown in Fig.\ \ref{fig:regimemap} which shows a two dimensional plot with Paschen parameters, $pd_{min}$ on x axis and $V_{min}$ on y axis and colored by the difference between $q_{eqbm}$ and $q_t$. We use the baseline parameters for particles from tables \ref{tab:tbparams}, \ref{tab:plasmaparams}, and \ref{tab:multipartparams} while only varying the Paschen breakdown parameters in Fig.\ \ref{fig:regimemap}. Two cases with $V_c=1.54\,\mathrm{V}$ (baseline case with Teflon particles contacting aluminium) and another case of $V_c=1\,\mathrm{V}$ (similar to polyethylene terephthalate (PET) particles and aluminium) are shown in Fig.\ \ref{fig:regimemap}(a) and (b), respectively. The line in Fig.\ \ref{fig:regimemap}(a) and (b) separates plasma feasible region (to the right) from the infeasible region to its left. Fig.\ \ref{fig:regimemap}(a) also marks the Paschen parameters for some common gases and indicates that gases like $\mathrm{CO_2}$ and $\mathrm{SO_2}$ are in the infeasible region with teflon-aluminium contact charging. With $V_c= 1\,\mathrm{V}$, the conditions are more stringent with infeasibility extending to $\mathrm{O_2}$ and $\mathrm{N_2O}$ environments. 

The analysis performed in this section therefore provides guidelines for selecting contact materials when targeting specific plasma-assisted conversion pathways. It should be noted that $q_{eqbm}$ also depends on particle radius, channel width and solids fraction, which can be modified to enable plasmas in gases with higher breakdown thresholds.

%% file: granular.tex
\label{sec:granular}
In this section, we model a realistic granular system where particles have a distribution of velocities. We solve the coupled-tribocharging-plasma model equations used in the previous sections (sections \ref{sec:singlepart} and \ref{sec:collectpart1}) with the additional effect of velocity distribution functions characterized by a granular temperature. We address here the questions regarding how overall plasma species production vary with granular temperature and solids fraction in a  realistic reactor, such as a vibrating bed used in previous tribocharging studies \cite{liu2022effect}. The modified tribocharging-plasma model is first derived in this section using velocity distributions and a numerical solve is performed using discrete velocity groups. Distribution function averaged quantities are then calculated and sensitivity studies are performed to characterize the impact of granular temperature and solids fraction on overall plasma species production.

\begin{figure}
    \centering
    \includegraphics[width=0.99\linewidth]{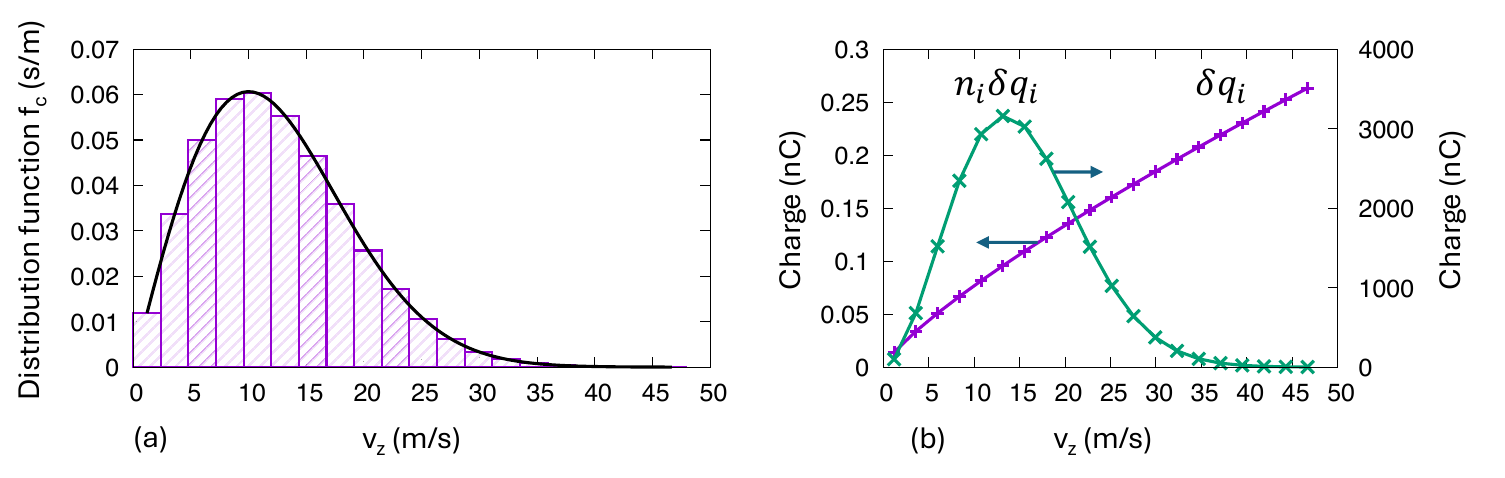}
    \caption{(a) Colliding velocity distribution function (black) with 20 discrete velocity groups (bars) at a granular temperature of 100 $\mathrm{m^2/s^2}$ and (b) tribocharge gained by each velocity group (purple) along with group number density weighted tribocharge (green).}
    \label{fig:vgroup}
\end{figure}

Let us consider a multi particle system with a distribution of particle velocities in a one dimensional system as shown in Fig.\ \ref{fig:spacecharge}. The velocity distribution is characterized by a granular temperature $\theta$ that depends on individual particle velocities ($v_p$) given by:
\begin{eqnarray}
    \theta=\frac{1}{3}\sum_p v_p^2
\end{eqnarray}
We assume that there is no bulk motion of particles and the average velocity over all particles sum to 0 (i.e. $\sum_p v_p=0$). The velocity distribution function associated with this system particles is given by:
\begin{eqnarray}
    f(v_x, v_y, v_z)= \left(\frac{1}{2\pi \theta}\right)^{3/2} \exp(-(v_x^2+v_y^2+v_z^2)/2\theta) \label{eq:distfunc}
\end{eqnarray}
where $v_x$, $v_y$, and $v_z$ are the 3 components of particle velocity along $x$, $y$, and $z$ directions, respectively. The one dimensional distribution function (say function of $v_z$ alone) for particles colliding with the walls is different from Eq.\ \ref{eq:distfunc} and can be derived by considering the flux on the surface with normal along the z direction as:
\begin{eqnarray}
    f_c(v_z)=\frac{\displaystyle \int_{-\infty}^{\infty} \int_{-\infty}^{\infty}  v_z f(v_x,v_y,v_z) dv_x dv_y}{\displaystyle \int_{0}^{\infty} \int_{-\infty}^{\infty} \int_{-\infty}^{\infty}  v_z f(v_x,v_y,v_z) dv_x dv_y dv_z}\\
    f_c(v_z)=\frac{v_z}{\theta} exp(-v_z^2/2\theta)
\end{eqnarray}
We use a velocity group based approach where we sample groups of velocities from the $f_c(v_z)$ as shown in Fig.\ \ref{fig:vgroup}(a) and apply the same equations from section \ref{sec:collectpart1}.
Each velocity group $i$ from Fig.\ \ref{fig:vgroup}(a) approaches the wall with a different velocity; the contact charge achieved by each group will be different. The bulk potential from Eq.\ \ref{eq:channelpot} will now include a weighted sum of charge possessed by each velocity group $i$ as:
\begin{eqnarray}
    V_{b}(z)=\frac{\sum_i n_i q_i}{2\epsilon_0}(t^2-z^2) \label{eq:vbg}
\end{eqnarray}
where $n_i$ and $q_i$ are the number density and charge associated with each velocity group.
The solution for $q_{t}$ and $z_t$, the particle charge and tangential location with the Paschen curve, will need to be solved using a set of coupled non-linear equations:
\begin{eqnarray}
     \frac{dV_{img}}{dz}(q_{ti},z_{ti}) + \frac{dV_b}{dz}(q_{ti},z_{ti})=
    \frac{d B_{p}}{dz}(z_{ti}) \label{eq:tangent5}\\
    V_{img}(q_{ti},z_{ti})+V_{b}(q_{t1},q_{t2}...,q_{tn},z_{ti})=B_p(z_{ti}) \label{eq:tangent6}
\end{eqnarray}
Eqs.\ \ref{eq:tangent5} and \ref{eq:tangent6} are a set of $2m$ non-linear equations with $2m$ variables ($q_{ti}$ and $z_{ti}$) where $m$ is the number of velocity groups. We have solved this equation set numerically and observed that the solutions remarkably converged to a condition where all tangential locations and charges are the same, i.e. $q_{t1}=q_{t2}=...q_{tm}=q_t$ and $z_{t1}=z_{t2}=...=z_{tm}=z_t$. If we make this assumption, Eqs.\ \ref{eq:tangent5} and \ref{eq:tangent6} will be identically satisfied by the solution to Eqs.\ \ref{eq:tangent3} and \ref{eq:tangent4}, after we simplify the $\sum_i n_i q_i$ from Eq.\ \ref{eq:vbg} to be just $n q_t$.
Once we have the tangential charge and locations, we continue the same procedure to integrate the plasma particle and energy balance equations (Eqs.\ \ref{eq:pb} and \ref{eq:eb2}) along with radical and excited species production for each velocity group. Fig.\ \ref{fig:vgroup}(b) shows the contact charge gained by each velocity group after collision when colliding with a charge of $q_t$. The contact charge varies directly with impact velocity although a number density weighted contact charge follows the velocity distribution function from Fig.\ \ref{fig:vgroup}(a).

\begin{figure}
    \centering
    \includegraphics[width=0.95\linewidth]{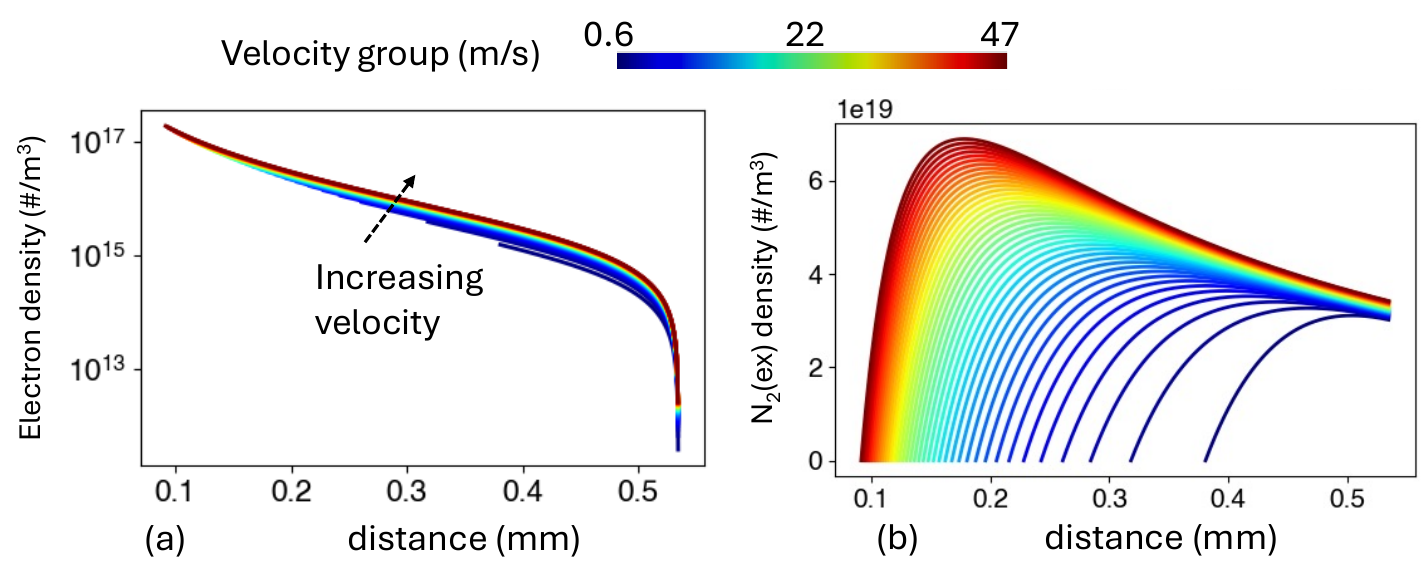}
    \caption{(a) electron number density and (b) $\mathrm{N_2(ex)}$ density as a function of distance traversed along the Paschen curve for 40 velocity groups ranging from 0.6 m/s to 47 m/s indicating increased plasma production and longer active discharge distance for higher velocities.}
    \label{fig:vdistdens}
\end{figure}

Fig.\ \ref{fig:vdistdens} shows an example calculation for a granular temperature of 100 $\mathrm{m^2/s^2}$ and a particle number density of 7.07e5 $\mathrm{\#/m^3}$ (solids volume fraction of 0.01) with 40 velocity groups indicating a distribution of electron and radical density for each velocity group. As expected, higher electron densities are achieved in the velocity groups with higher velocity. The charge relaxation onset is also at a closer point to the wall for higher velocities as is expected from higher post-contact charge. Figure \ref{fig:vdistdens}(b) shows the excited nitrogen density for various velocity groups with the highest velocity group with $\sim$ 47 m/s achieving the highest peak density. A similar trend is observe for N radical densities as well with peak densities similar in magnitude as shown in Fig.\ \ref{fig:multip_plasmaparams}(c). A steady decrease in the excited species density is observed in Fig.\ \ref{fig:vdistdens}(b) after achieving peak densities mainly arising from lower production rates from electron density reduction and an increase in discharge volume as the particle moves away from the wall. The second term on the right hand side of Eq.\ \ref{eq:ode3} that accounts for the change in discharge volume is driving this effect.

\begin{figure}
    \centering
    \includegraphics[width=0.99\linewidth]{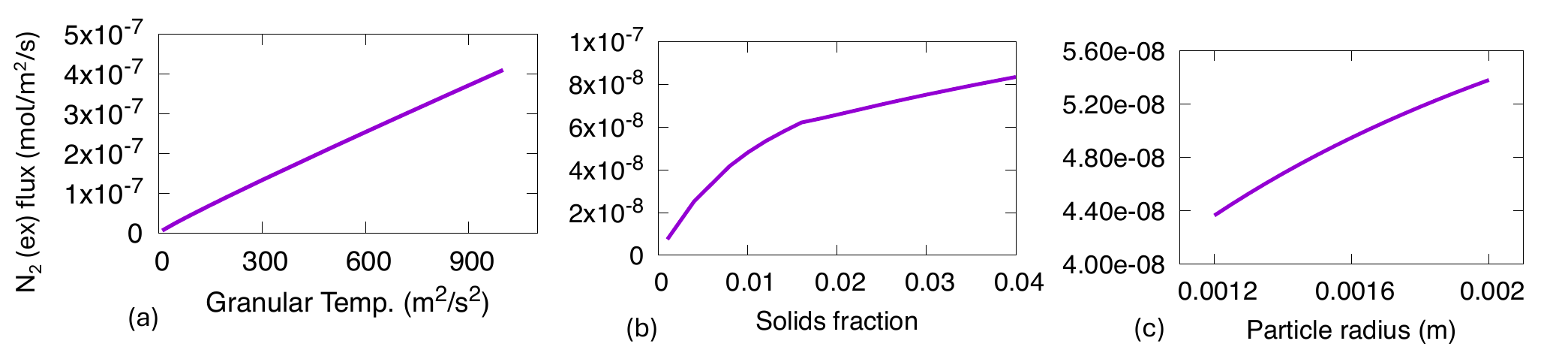}
    \caption{$\mathrm{N_2(ex)}$ flux from the wall as a function (a) granular temperature, (b) solids fraction and (c) particle radius.}
    \label{fig:n2exflux}
\end{figure}

The overall excited or dissociated species flux can be obtained by calculating the number of moles of species formed per collision and multiplying it with the particle flux on the surface. When using discrete velocity groups to approximate the distribution function, the flux of $\mathrm{N_2(ex)}$ species can be obtained as:
\begin{equation}
    \Gamma_{N_2(ex)}= \frac{n \bar{c}}{4} \int {\left( f_c n_{N_2(ex)} \Omega_{f} \right) dv_x}
\end{equation}
where $\bar{c}=\sqrt{\frac{8 \theta}{\pi}}$ is the average particle velocity and $\Omega_f$ is the final discharge volume for each particle ($\Omega_f = z_t A_{coll}$).
The variation of this overall flux is shown in Fig.\ \ref{fig:n2exflux} for the \ce{N2}(ex) indicating a direct dependence on granular temperature, solids fraction, and particle radius. The trend associated with granular temperature is similar to Fig.\ \ref{fig:multip_sens}(b) where higher particle velocities are directly correlated to density of excited species which is a direct consequence of increased contact charge acquired by the particle at higher velocities. 
\revise{The variation of overall excited species flux shows a non-linear increasing trend with solids fraction, as indicated by Fig.\ \ref{fig:n2exflux}(b). Although, an inverse trend was observed \ce{N2}(ex) density per particle in Fig.\ \ref{fig:multip_sens}(d). 
The main reason here is the increased particle number density from higher solids fraction is a dominating effect on \revise{\ce{N2}(ex) flux} compared to the reducing effect of increased space charge on \revise{\ce{N2}(ex) concentration}, as seen in Fig.\ \ref{fig:multip_sens}(d).} The excited species flux is positively correlated with particle radius similar to Fig.\ \ref{fig:multip_sens}(f) which is mainly arising from higher contact charge due to larger collisional area and lower impact of image potential.

This improved model that accounts for velocity distributions provides estimates for overall excited species fluxes that are critical intermediates for plasma catalytic conversion. Larger particles, greater solids fraction and higher granular temperature favor increased production of \ce{N2}(ex) species, thereby providing insights into future experiments and reactor designs.